\documentclass[review]{elsarticle}

\usepackage{lineno,hyperref,mathtools}
\usepackage[letterpaper,top=2cm,bottom=2cm,left=3cm,right=3cm,marginparwidth=1.75cm]{geometry}

\journal{Journal of \LaTeX\ Templates}

\usepackage{amssymb}
\usepackage{bm}
\usepackage{amssymb}
\usepackage{amsthm}
\usepackage{multirow,booktabs}
\usepackage{cases}
\usepackage{threeparttable}

\newcommand{\F}{\mathbb{F}}
\newcommand{\Z}{\mathbb{Z}}
\newcommand{\C}{\mathcal{C}}
\newcommand{\lcm}{\mathrm{lcm}}
\newcommand{\ord}{\mathrm{ord}}
\newcommand{\MinRep}{\mathrm{MinRep}}

\newtheorem{theorem}{Theorem}

\newtheorem{remark}{Remark}

\newtheorem{lemma}{Lemma}

\newtheorem{example}{Example}
\begin{document}
	
	\begin{frontmatter}
		
\title{Two types of narrow-sense negacyclic BCH codes}
\tnotetext[mytitlenote]{}

\author[mymainaddress]{Yanhui Zhang}
\ead{zhangyanhui0327@163.com}

\author[mymainaddress]{Li Liu\corref{mycorrespondingauthor}}
\cortext[mycorrespondingauthor]{Corresponding author}
\ead{liuli-1128@163.com}

\address[mymainaddress]{School of Mathematics, Hefei University of Technology, Hefei 230009, China}
\author[myaddress]{Xianhong Xie}
\ead{xianhxie@ahau.edu.cn}
\address[myaddress]{School of Information and Computer Science, Anhui Agricultural University, Hefei 230036, Anhui, China}	
\begin{abstract}
Negacyclic BCH codes are an important subclass of negacyclic codes and are the best linear codes in most cases, but their parameters are hard to determine. In this paper, we mainly study two types of negacyclic BCH codes of length $n=\frac{q^{m}-1}{4},\frac{q^{m}+1}{4}$, and give their dimensions and the lower bound on their minimum distance. Furthermore, we provide the weight distribution of narrow-sense neagcyclic BCH codes of length $n=\frac{q^m-1}{4}$ for some special designed distances.

\end{abstract}
		
 \end{frontmatter}
\section{Introduction}
Throughout this paper, let $q$ be an odd prime power and $\F_{q}$ be a finite field of size $q$. An $[n,k]$ linear code $\C$ over $\F_q$ is a $k$-dimension linear subspace of $\F_q^{n}$. Let $\lambda\in \F_{q}^{*}$, if for any $(c_{0}, c_{1},\ldots, c_{n-1}) \in \C$ implies $(\lambda c_{n-1}, c_{0}, c_{1},\ldots, c_{n-2})\in \C$, then $\C$ is called a $\lambda-$constacyclic code. Let $R=\F_{q}[x]/(x^n-\lambda)$ be a residue class ring, identifying any vector $(c_{0}, c_{1},\ldots, c_{n-1})\in \F_q^{n}$ with
\begin{center}
$c_{0}+c_{1}x+c_{2}x^{2}+\cdots+c_{n-1}x^{n-1}\in \F_q[x]/ ( x^{n}-\lambda)$,
\end{center}
then a $\lambda-$constacyclic code $\C$ over $\F_q$ corresponds to an ideal of $R$. Clearly, $R$ is a principal ideal ring, then $\C$=$\langle g(x) \rangle$ for any $\lambda-$constacyclic code $\C$, where $g(x)$ is the smallest degree monic polynomial satisfying $g(x)|(x^{n}-\lambda)$. Furthermore, $g(x)$ and $h(x) = \frac{x^{n}-\lambda}{g(x)}$ are referred to as the generator polynomial and check polynomial of $\C$, respectively. If $\lambda=-1$, $\C$ is called a negacyclic code, while $\C$ is called a cyclic code for $\lambda=1$.
	
Suppose $n\in\Z^{+}$ with $\gcd(n,q)=1$, $m=\ord_{2n}(q)$ is the order of $q$ modulo $2n$, where $\Z^{+}$ is the set of all positive integers. Put $\F_{q^m}^*=\langle\alpha\rangle$ and $\beta= \alpha^{\frac{q^{m}-1}{2n}}$, then $\beta$ is a primitive $2n$-th root of unity in $\F_{q^m}$. Denote the minimal polynomial of $\beta ^{1+2i}$ over $\F_q$ as $M_{\beta^{{1+2i}}}(x)$ for any $0\leq i\leq n-1$. For $\delta,b\in\Z^{+}$, define
\begin{center}
$g_{(n,q,\delta,b)}(x)=\lcm \left(M_{\beta^{{1+2b}}}(x),M_{\beta^{{1+2(b+1)}}}(x),\ldots,M_{\beta^{{1+2(b+\delta-2)}}}(x)\right)$,
\end{center}
where $\lcm$ represents the least common multiple of these polynomials and $2\leq \delta \leq n$. Let $\C_{(n,q,\delta,b)}=$$\langle g_{(n,q,\delta,b)} \rangle$, then $\C_{(n,q,\delta,b)}$ is referred to as a negacyclic BCH code, where $\delta$ is called the designed distance of $\C$. If $b=0$, it is called a narrow-sense negacyclic BCH (NS NBCH) code and is denoted as $\C_{(n,q,\delta)}$, otherwise it is a non-narrow-sense negacyclic BCH (NNS NBCH) code.

BCH codes were first introduced by Hocquenghem in \cite{RefJ10}. Gorenstein and Zierler extended it to general finite fields in \cite{RefJ12}. In the last few decades, the cyclic codes were widely investigated in \cite{RefJ1}, \cite{RefJ2}, \cite{RefJ5}-\cite{RefJ9}, \cite{RefJ14} and \cite{RefJ18}-\cite{RefJ32}. Recently, many researchers have developed a strong interest in the study of negacyclic codes.
	
Negacyclic codes were proposed by Berkelamp in \cite{RefJ3}, \cite{RefJ4}. Krishna and Sarwate found that optimal linear codes can be constructed from negacyclic codes in \cite{RefJ15}. Recently, Kai et al. constructed many optimal quantum codes from negacyclic BCH codes in \cite{RefJ16}, \cite{RefJ17}. At the same time, Zhu et al. obtained some quantum codes with good parameters from negacyclic BCH codes of length $n=\frac{q^{2m}-1}{q-1}$ in \cite{RefJ28}.  In \cite{RefJ21}, Pang et al. studied the parameters of three classes of negacyclic BCH codes. In \cite{RefJ13}, Guo et al. studied a class of $q^2$-ary NS NBCH and NNS NBCH codes of length $n=\frac{q^{2m}-1}{2}$. In \cite{RefJ23}, Wang et al. investigated two families of negacyclic BCH codes of length $n=\frac{q^{m}-1}{2}$, $\frac{q^{m}+1}{2}$ and disscussed their parameters. 
	
Building upon the aforementioned work, we construct two classes of negacyclic BCH codes of length $n=\frac{q^{m}-1}{4}$, $\frac{q^{m}+1}{4}$, and completely determine their parameters, showing that they contain some optimal codes to the Sphere packing bound. The structure of this article is outlined as follows: In Section 2, we provide some preliminaries. In Section 3, we study the parameters of the narrow-sense negacyclic BCH code $\C_{(\frac{q^m-1}{4},q,\delta)}$, and give the minimum distance for $\delta=2,3$. Additionally, we determine the weight distribution of $\C_{(\frac{q^m-1}{4},q,\delta)}$ for $\delta$ in some special ranges. In Section 4, we study the parameters of the code $\C_{(\frac{q^m+1}{4},q,\delta)}$, and investigate the minimum distance for $\delta=2$. Finally, Section 5 concludes this paper.
\section{Preliminaries}
In this section, we present fundamental concepts and established findings about BCH codes.
\subsection{Basic notations}
For any $s$ with $0\leq s\leq n-1$, the $q$-cyclotomic coset of representative $s$ modulo $n$ is referred to as
\[C_{s}^{n}=\{sq^k\pmod{n}: \ 0\leq k\leq l_s-1\},\]
where $l_s$ is the smallest integer such that $s\equiv sq^{l_s}\pmod n$ and $l_s=|C^{n}_{s}|$. Let $CL(s)=\min\{i:\ i\in C_{s}^{n}\}$ be called the coset leader of $C_{s}^{n}$ and $\MinRep_{n}=\{CL(s): \ 0\leq s\leq n-1\}$. Furthermore, we denote $sq^k\pmod{n}$ as $[sq^{k}]_{n}$ for any $0\leq k\leq l_s-1$.
	
For $n\in\Z^{+}$ satisfying $\gcd(n,q)=1$, let $\F_{q^m}^*=\langle\alpha\rangle$ and $m=\ord_n(q)$. Denote $\gamma= \alpha^{\frac{q^{m}-1}{n}}$, then $\gamma$ is a primitive $n$-th root of unity in $\F_{q^m}$. Define	
\begin{center}
$g^{'}(x)=\lcm \left(m_{\gamma^{1}}(x),m_{\gamma^{2}}(x),\ldots,m_{\gamma^{\delta-1}}(x)\right)$,
\end{center}where $2\leq \delta \leq n$ and $m_{\gamma^{i}}(x)$ is the minimal polynomial of $\gamma ^{i}$ over $\F_q$. Denote $\C^{'}_{(n,q,\delta,1)}=\langle g^{'}(x)\rangle$,
then $\C^{'}_{(n,q,\delta,1)}$ is called a narrow-sense cyclic BCH code. Furthermore, d$(\C)$ and $\dim(\C)$ are denoted as the minimum distance and dimension of the code $\C$, respectively. Let $\underbrace{A,\ldots}_{i}=\underbrace{A,A,\ldots,A}_{i\ times}$ and $[u,v]=\{i:i\in \Z \ and\ u\leq i\leq v\}$.
\subsection{Known results}
We are listing some established conclusions, this will be required in Section 3, 4.
\begin{lemma}(\cite{RefJ15})\label{Le1}
Let $g(x)$ be a generator polynomial of negacyclic code $\C$ of length $n$ and $\beta$ be a primitive $2n$-th root of unity. If there exist $k,b,\delta\in\Z$ such that $\gcd(k,n)=1$, $2\leq \delta\leq n$ and
$$g(\beta^{1+2kb})=g(\beta^{1+2k(b+1)})=\cdots=g(\beta^{1+2k(b+\delta-2)})=0,$$
then d$(\C)\geq \delta$.
\end{lemma}
\begin{lemma}(\cite{RefJ9}\cite{RefJ30})\label{Le2}
(Sphere Packing Bound) Let $\C$ be a $q$-ary $[n,k,d]$ code, then 
$$q^{n-k}\geq \sum_{i=0}^{\lfloor\frac{d-1}{2}\rfloor}\begin{pmatrix}
n\\
i
\end{pmatrix}(q-1)^i.$$
If $2\mid d$, then
$$ q^{n-1-k}\geq \sum_{i=0}^{\frac{d-2}{2}}\begin{pmatrix}
n-1\\
i
\end{pmatrix}(q-1)^i.$$
\end{lemma}
\begin{lemma}(\cite{RefJ27})\label{Le3}
Let $m=2h\geq 4$, $\lambda|(q-1)$ and  $n=\frac{q^{m}-1}{\lambda}$.
\begin{itemize}
\item[(1)] Let $a,b,c\in\Z$, $f(a,b,c)=aq^{h}+\frac{b(q^{h}-1)}{\lambda}+c$, and 
	$\Delta_1=\{f(a,0,c):1\leq c<a\leq \frac{q-1}{\lambda}\},$
	$\Delta_2=\{f(a,b,c):1\leq c\leq a<\frac{q-1}{\lambda},1\leq b<\lambda\},$
	$\Delta_3=\{f(a,b,a+1):0\leq a<\frac{q-1}{\lambda},\frac{\lambda}{2}<b<\lambda\}.$	
If $1\leq i\leq \frac{q^{h+1}-1}{\lambda}$ satisfying $i\not\equiv0\pmod{q}$ and  $i\notin \Delta_1\cup \Delta_2\cup \Delta_3$, then $i\in \MinRep_{n}$.
\item[(2)] If $2\mid\lambda$ and $1\leq i\leq \frac{q^{h+1}-1}{\lambda}$, let $\Delta=\{\frac{c(q^h+1)}{2}:1\leq c\leq \frac{2(q-1)}{\lambda}\}$. Then
$$|C_{i}^{n}|=\begin{cases}
h,& \ \text{if}\ i\in \Delta;\\
m,&\ \text{otherwise}.
\end{cases}$$ 
\end{itemize}
\end{lemma}
\begin{lemma}(\cite{RefJ22})\label{Le4}
Let $n=\frac{q^m-1}{2}$ and $1\leq i\leq \lfloor\frac{m+6}{4}\rfloor$, then the $i-$th largest coset leader modulo $n$ is
$$\delta^{'}_{i}=\frac{q^m-q^{m-1}-q^{\lfloor\frac{m-3}{2}+i\rfloor}-1}{2}.$$
Moreover, \begin{center}
$|C_{\delta_{i}^{'}}^{n}|=$$\begin{cases}
\frac{m}{2},&\ \text{if}\ 2\mid m\ and\ i=1;\\
m,&\ \text{otherwise}.
\end{cases}$
\end{center}
\end{lemma}
\begin{lemma}(\cite{RefJ25},\cite{RefJ26})\label{Le5}
Let $m=2h+1\geq 5$ and $n=\frac{q^{m}+1}{2}$.
\begin{itemize}
\item[(1)]Let $a,b,c\in\Z$,  $f(a,b,c)=\frac{q^{h+1}+1}{2}+a\frac{q^{h}-1}{2}+bq^{h}+c$ and $g(a,b,c)=\frac{q^{h+1}-1}{2}+a\frac{q^{h}+1}{2}+bq^{h}+c$. Define 
\begin{center}
$\begin{cases}
X_{1}=\lbrace f(0,0,c): -\frac{q}{2} \leq c\leq \frac{q-2}{2}  \rbrace;\\
X_{2}=\lbrace f(0,b,0):1\leq b\leq \frac{q-1}{2} \rbrace;\\	
X_{3}=\lbrace f(2,b,0):0 \leq b\leq \frac{q-3}{2}\rbrace
\end{cases}$for $2\mid h$,
$\begin{cases}
X_{1}=\lbrace g(0,0,c):-\frac{q-2}{2}\leq c\leq \frac{q}{2}  \rbrace;\\		
X_{2}=\lbrace g(0,b,0):1\leq b\leq \frac{q-1}{2}  \rbrace;\\
X_{3}=\lbrace g(2,b,0):0 \leq b\leq \frac{q-3}{2} \rbrace
\end{cases}$for $2\nmid h$.
\end{center}If $1\leq i< \frac{2q^{h+1}-2q+1}{2}$ satisfying $i\not\equiv0\pmod{q}$ and $i\notin X_{1}\cup X_{2}\cup X_{3}$, then $i\in \MinRep_{n}$ and $|C_{i}^{n}| =2m$.
\item[(2)]If $q\equiv3\pmod{4}$, then the first three largest coset leaders modulo $n$ are: $$\delta_{1}^{'}=\frac{q^{m}+1}{4},\ \delta_{2}^{'}=\frac{q^{m}-1}{4}-\frac{q^{m-1}}{2},\ 
\delta_{3}^{'}=\frac{q^{m}+1}{4}-\frac{q^{m-1}+q}{2}.$$ 
Moreover, $|C_{\delta_{1}^{'}}^{n}|=1$ and $|C_{\delta_{2}^{'}}^{n}|=|C_{\delta_{3}^{'}}^{n}|=2m$.
\end{itemize}
\end{lemma}
\begin{lemma}(\cite{RefJ24},\cite{RefJ29})\label{Le6}
Let $\lambda\mid (q+1)$ and $n=\frac{q^{m}+1}{\lambda}$.
\begin{itemize}
\item[(1)]If $\lambda=1$, let $k,h,l\in \Z$ such that $1\leq k< m$, 	$-\frac{l(q^{m-k}-1)}{q^{k}+1}<h<\frac{l(q^{m-k}+1)}{q^{k}-1}$ and $1\leq l\leq \frac{q^{k}-1}{2}$. Then $a\in \MinRep_{n}$ with $0\leq a\leq q^{m}$ if and only if $\frac{q^{m}+1}{2}\geq a$ and $a\neq lq^{m-k}+h$.
\item[(2)]If $\lambda\neq1$ and $1\leq i\leq n-1$, then $i\in \MinRep_{n}$ if and only if $\lambda i\in \MinRep_{q^m+1}$. Moreover, $|C_{i}^{n}|=|C_{\lambda i}^{\lambda n}|$.
\end{itemize}
\end{lemma}
\section{The case of $n=\frac{q^m-1}{4}$}
In this section, we consistently assume $q\equiv 1\pmod4$ or $q\equiv 3\pmod4$ and $2\mid m$ when $n=\frac{q^m-1}{4}$. We first investigate the parameters of the code $\C_{(n,q,\delta)}$ with large dimension. 
\begin{lemma}\label{Le7}
Let $m=2h\geq 4$ and $n=\frac{q^m-1}{2}$. 
\begin{itemize} 
\item If $q\equiv1\pmod4$ or $q\equiv3\pmod4 \ \text{and} \ 4\mid m$, define
$$\begin{aligned}
&T_1=\{(2u+1)q^h+2v,(2u+1)q^h+2v+\frac{q^h-1}{2}:1\leq v\leq u<\frac{q-3}{4}\},\\
&T_2=\{2uq^h+2v+1,2uq^h+2v+\frac{q^h+1}{2}:0\leq v< u<\frac{q-1}{4}\},\\
&T_3=\begin{cases}
\{\frac{(q-1)q^h}{2}+2v+1:0\leq v< \frac{q-1}{4}\},&\ \text{if}\ q\equiv1\pmod4;\\
\{\frac{(q-1)q^h}{2}+2v:1\leq v\leq \frac{q-3}{4}\},&\ \text{if}\ q\equiv3\pmod4 \ \text{and}\ m\equiv0 \pmod4.
\end{cases}
\end{aligned}$$
\item If $q\equiv3\pmod4$ and $m\equiv2\pmod 4$, define
$$\begin{aligned}
&T_1=\{(2u+1)q^h+2v+\frac{q^h+1}{2}:0\leq v\leq u<\frac{q-3}{4}\},\\
&T_2=\{2uq^h+2v+1:0\leq v< u\leq \frac{q-3}{4}\},\\
&T_3=\{(2u+1)q^h+2v,2uq^h+2v+\frac{q^h-1}{2}:1\leq v\leq u\leq\frac{q-3}{4}\}.
\end{aligned}$$
\end{itemize}
If $1\leq i\leq \frac{q^{h+1}-1}{2}$ is odd satisfying $i\not\equiv0\pmod{q}$ and $i\notin T_1\cup T_2\cup T_3$, then $i\in \MinRep_{n}$.
\begin{proof}
Let $1\leq i\leq \frac{q^{h+1}-1}{2}$ satisfying $i\not\equiv0\pmod{q}$, then $i\notin \MinRep_{n}$ only when
\begin{equation}\label{e1}
\begin{cases}
i=aq^h+c,&\ 1\leq c<a\leq \frac{q-1}{2};\\
i=aq^h+c+\frac{q^h-1}{2},&\ 1\leq c\leq a<\frac{q-1}{2}.
\end{cases}
\end{equation}by Lemma \ref{Le3}.
\\$\bf{Case\ 1.}$ If $q\equiv1\pmod 4$ or $q\equiv3\pmod4$ and $4\mid m$, then $2\mid \frac{q^{h}-1}{2}$, i.e., we only need to consider the case of $2\nmid (aq^h+c)$ by (\ref{e1}).
\begin{itemize}
\item[1)]
If $2\nmid a$, then $2\mid c$. We can assume that $a=2u+1$ and $c=2v$, then we have 
\begin{center}
$\begin{cases}
1\leq 2v<2u+1\leq \frac{q-1}{2};\\
1\leq 2v\leq 2u+1<\frac{q-1}{2},
\end{cases}$ $\Rightarrow$
$\begin{cases}
	1\leq v\leq u\leq \frac{q-3}{4};\\
	1\leq v\leq u<\frac{q-3}{4}
\end{cases}$
\end{center} by (\ref{e1}), where $u$, $v$ are integers. Thus $q\equiv3\pmod 4$ and $i\notin \MinRep_{n}$ is odd when $i\in T_1\cup T_3$, $q\equiv1\pmod 4$ and $i\notin \MinRep_{n}$ is odd when $i\in T_1$, where $$\begin{cases}
T_1=\{(2u+1)q^h+2v,(2u+1)q^h+2v+\frac{q^h-1}{2}:1\leq v\leq u<\frac{q-3}{4}\};\\
T_3=\{\frac{(q-1)q^h}{2}+2v:1\leq v\leq \frac{q-3}{4}\}.
\end{cases}$$
\item[2)]
If $2\mid a$, then $2\nmid c$. We can assume that $a=2u$ and $c=2v+1$, then we have 
\begin{center}
$\begin{cases}
1\leq 2v+1<2u\leq \frac{q-1}{2};\\
1\leq 2v+1\leq 2u<\frac{q-1}{2},
\end{cases}$ $\Rightarrow$
$\begin{cases}
0\leq v< u\leq \frac{q-1}{4};\\
0\leq v< u<\frac{q-1}{4}
\end{cases}$
\end{center} by (\ref{e1}), where $u$, $v$ are integers. Thus $q\equiv3\pmod 4$ and $i\notin \MinRep_{n}$ is odd when $i\in T_2$, $q\equiv1\pmod 4$ and $i\notin \MinRep_{n}$ is odd when $i\in T_2\cup T_3$, where $$\begin{cases}
T_2=\{2uq^h+2v+1,2uq^h+2v+\frac{q^h+1}{2}:0\leq v< u<\frac{q-1}{4}\};\\
T_3=\{\frac{(q-1)q^h}{2}+2v+1:0\leq v< \frac{q-1}{4}\}.
\end{cases}$$
\end{itemize}
$\bf{Case\ 2.}$ If $q\equiv3$ and $m\equiv2\pmod 4$, then $2\nmid \frac{q^{h}-1}{2}$. We obtain the results using the same method as in Case 1.
This concludes our proof.
\end{proof}
\end{lemma}
Next we will give the dimension of the code $\C_{(n,q,\delta)}$ with $2\leq \delta\leq \frac{q^{\frac{m+2}{2}}+5}{4}$, where $m=2h\geq 4$.
\begin{theorem}\label{t1}
Let $q\equiv1\pmod4$ or $q\equiv3\pmod4$ and $4\mid m$, $m=2h\geq 4$ and $n=\frac{q^m-1}{4}$. Then the code $\C_{(n,q,\delta)}$ has parameters $[n,k,d]$, where \begin{center}
$\begin{cases}
d\geq\delta+1,& \ \text{if}\ \delta\equiv\frac{q+1}{2}\pmod q;\\
d\geq\delta,&\  \text{otherwise}
\end{cases}$ 
\end{center}and the dimension $k$ is provided as follows:
\begin{itemize}
\item[(1)]If $q\geq 5$ and $2\leq\delta\leq q^{h}+1$, then 
$$k=\begin{cases}
n-m\big\lceil \frac{(2\delta-3)(q-1)}{2q}\big\rceil,& \text{if}\ 2\leq \delta\leq \frac{q^h-1}{4}+1;\\
n-m\big\lceil \frac{(2\delta-3)(q-1)}{2q}\big\rceil+h,& \text{if}\ \frac{q^h-1}{4}+2\leq \delta\leq\frac{3(q^h+3)}{4}-1;\\
n-m\big\lceil \frac{(2\delta-3)(q-1)}{2q}\big\rceil+m,& \text{if}\ \frac{3(q^h+3)}{4}\leq \delta\leq q^h+1.
\end{cases}$$
\item[(2)]If $q=5$ and $q^h+2\leq \delta\leq\frac{q^{h+1}+5}{4}$, then $$k=n-m\big\lceil \frac{(2\delta-3)(q-1)}{2q}\big\rceil+2m.$$
\item[(3)]If $q>5$, let $\tau\in\Z$ with $1\leq \tau<\frac{q-3}{4}$.
\begin{itemize}	
\item[3.1)]If $q^h+2\leq \delta\leq(\tau+1) q^h+1$, then
$$k=\begin{cases}
n-m\big(\big\lceil \frac{(2\delta-3)(q-1)}{2q}\big\rceil-2\tau^2\big),& \text{if}\ \tau (q^h+1)+1\leq \delta\leq \tau q^h+\frac{q^h+3}{4};\\
n-m\big(\big\lceil \frac{(2\delta-3)(q-1)}{2q}\big\rceil-(2\tau ^2+\tau)\big)+h,& \text{if}\ \tau (q^h+1)+\frac{q^h+7}{4}\leq \delta\leq \tau q^h+\frac{q^h+3}{2};\\
n-m\big(\big\lceil \frac{(2\delta-3)(q-1)}{2q}\big\rceil-2(\tau ^2+\tau)\big)+h,& \text{if}\ \tau (q^h+1)+\frac{q^h+3}{2}\leq \delta\leq \tau q^h+\frac{3q^h+5}{4};\\
n-m\big(\big\lceil \frac{(2\delta-3)(q-1)}{2q}\big\rceil-(2\tau ^2+3\tau +1)\big),& \text{if}\ \tau (q^h+1)+\frac{3q^h+1}{2}+2\leq \delta\leq (\tau+1) q^h+1.
\end{cases}$$	
\item[3.2)]If $q\equiv1\pmod4$ and $\frac{q-1}{4}q^h+2\leq\delta\leq \frac{q^{h+1}+5}{4}$, then
$$k=\begin{cases}
n-m\big(\big\lceil \frac{(2\delta-3)(q-1)}{2q}\big\rceil-(\delta-\frac{q-1}{4}q^h+\frac{(q+1)(q-5)}{8})\big),& \text{if}\ \frac{q-1}{4}q^h+2\leq \delta\leq\frac{q-1}{4}(q^h+1)+1;\\
n-m\big(\big\lceil \frac{(2\delta-3)(q-1)}{2q}\big\rceil-\frac{(q-1)^2}{8}\big),& \text{if}\ \frac{q-1}{4}(q^h+1)+2\leq \delta\leq \frac{q^{h+1}+5}{4}.
\end{cases}$$
\item[3.3)]If $q\equiv3\pmod4$ with $4\mid m$ and $ \frac{(q-3)q^h+q+1}{4}\leq \delta\leq \frac{q^{h+1}+5}{4}$, then
$$k=\begin{cases}
n-m\big(\big\lceil \frac{(2\delta-3)(q-1)}{2q}\big\rceil-\frac{(q-3)^2}{8}\big),& \text{if}\ \frac{q^{h+1}-3q^h+q+1}{4}\leq \delta\leq\frac{q^{h+1}-2q^h+3}{4};\\
n-m\big(\big\lceil \frac{(2\delta-3)(q-1)}{2q}\big\rceil-\frac{(q-1)(q-3)}{8}\big)+h,& \text{if}\ \frac{q^{h+1}-2q^h+q}{4}+1\leq \delta\leq \frac{q^{h+1}-q^h+2}{4}+1;\\
n-m\big(\big\lceil \frac{(2\delta-3)(q-1)}{2q}\big\rceil-\frac{(q-1)^2}{8}\big),& \text{if}\ \frac{q^{h+1}-q^h+q+3}{4}\leq \delta\leq \frac{q^{h+1}+5}{4}.
\end{cases}$$	
\end{itemize}
\end{itemize}
\begin{proof}
It is clear that the generator polynomal of $\C_{(n,q,\delta)}$ is $g(x)=\lcm(M_{\beta^{1}}(x),M_{\beta^{3}}(x),\ldots,M_{\beta^{1+2(\delta-2)}}(x))$,  where $\beta$ is a primitive $2n$-th root of unity. Let $\Gamma=\lbrace 1+2i:0\leq i\leq \delta-2,\ \ 1+2i\not\equiv0\pmod q\rbrace$, where $2\leq \delta\leq\frac{q^{h+1}+5}{4}$. Then $1+2i\in \Gamma$ is a coset leader modulo $2n$  except $1+2i\in\cup^{3}_{i=1} T_{i}$ by Lemma \ref{Le7}, where $T_i$ is defined in Lemma \ref{Le7}. Define $T_0=\{\frac{(2t+1)(q^h+1)}{2}:0\leq t\leq \frac{q-2}{2}\}$, then
$$|C_{1+2i}^{2n}|=\begin{cases}
h,& \text{if}\ 1+2i\in T_0;\\
m,& \text{if}\ 1+2i\in \Gamma\textbackslash T_0
\end{cases}$$ by Lemma \ref{Le3}.
Note that $T_i\cap T_j=\emptyset$ for $0\leq i\neq j\leq 3$, the dimension of the code $\C_{(n,q,\delta)}$ is 
\begin{equation}\label{eq2}
k=n-m|\Gamma|+m\sum_{i=1}^{3}|\Gamma\cap T_i|+\frac{m}{2}|\Gamma\cap T_0|.
\end{equation}
Note that $\min\{T_0\}=\frac{q^h+1}{2}$, $\min\{T_1,T_2,T_3\}=2q^h+1$ and
$$\begin{aligned}
|\Gamma|&=\delta-1-|\{1+2i:0\leq i\leq \delta-2,\ \ 1+2i\equiv0\pmod q\}|\\
&=\delta-1-(\big\lfloor \frac{2\delta-3-q}{2q}\big\rfloor+1)=\big\lceil \frac{(2\delta-3)(q-1)}{2q}\big\rceil.
\end{aligned}$$ 
			
If $q\geq 5$ and $2\leq\delta\leq q^{h}+1$, we have the following. 
\\For $2\leq \delta\leq \frac{q^h-1}{4}+1$, we have $\frac{q^h-1}{2}-1\geq 2\delta-3$. We have $\Gamma\cap T_i=\emptyset$ for any $i\in [0,3]$, it follows from (\ref{eq2}) that $$k=n-m\big\lceil \frac{(2\delta-3)(q-1)}{2q}\big\rceil.$$
Similarly, we have $\Gamma\cap T_0=\{\frac{q^h+1}{2}\}$ and $\Gamma\cap (\cup _{i=1}^{3}T_i)=\emptyset$ when $\frac{q^h-1}{4}+2\leq \delta\leq \frac{3(q^h+3)}{4}-1$, $\Gamma\cap T_0=\{\frac{q^h+1}{2},\frac{3(q^h+1)}{2}\}$ and $\Gamma\cap (\cup _{i=1}^{3}T_i)=\emptyset$ when $\frac{3(q^h+3)}{4}\leq \delta\leq q^h+1$, then we can get the results.
			
If $q=5$ and $q^h+2\leq \delta\leq\frac{q^{h+1}+5}{4}$, we have $\Gamma\cap T_0=\{\frac{q^{h}+1}{2},\frac{3(q^h+1)}{2}\}$ and $\Gamma\cap (\cup _{i=1}^{3}T_i)=\{2\cdot q^h+1\}$. It follows from (\ref{eq2}) that $$k=n-m\big\lceil \frac{(2\delta-3)(q-1)}{2q}\big\rceil+2m.$$
			
If $q> 5$ and $q^h+2\leq \delta\leq \frac{q^{h+1}+5}{4}$, we only prove the case of $\tau (q^h+1)+1\leq \delta\leq \tau q^h+\frac{q^h+3}{4}$, as other cases are similar. Note that $2\tau q^h+2\tau -1\leq 2\delta-3\leq 2\tau q^h+\frac{q^h-3}{2}$, then 
\begin{center}
$\Gamma\cap T_{3}=\emptyset$, $\Gamma\cap T_{0}=\lbrace \frac{(4t+1)(q^{h}+1)}{2},\  \frac{(4t+3)(q^{h}+1)}{2} : 0\leq t<\tau \rbrace$,\\
$\Gamma\cap T_{1}=\{(2u+1)q^h+2v,(2u+1)q^h+2v+\frac{q^h-1}{2}:1\leq v\leq u<\tau\}$,\\
$\Gamma\cap T_{2}=\{2uq^h+2v+1,2uq^h+2v+\frac{q^h+1}{2}:0\leq v< u<\tau\}\cup \{2\tau q^h+2v+1:0\leq v<\tau\}$.	
\end{center}
It is clear that $| \Gamma\cap T_{3}|=0$, $|\Gamma\cap T_{0}|=2\tau$, $|\Gamma\cap T_{1}|=\tau(\tau-1)$ and $|\Gamma\cap T_{2}|=\tau^2$, it follows from (\ref{eq2}) that $$k=n-m\big(\big\lceil \frac{(2\delta-3)(q-1)}{2q}\big\rceil-2\tau^2\big).$$
This concludes our proof.
\end{proof}
\end{theorem}
\begin{theorem}\label{t2}
Let $q\equiv3\pmod4$, $m=2h\geq 4$ and $n=\frac{q^m-1}{4}$. Then the code $\C_{(n,q,\delta)}$ has parameters $[n,k,d]$, where
\begin{center}
$\begin{cases}
d\geq \delta+1, \ &\text{if}\ \delta\equiv\frac{q+1}{2}\pmod q;\\
d\geq \delta,\  &\text{otherwise}
\end{cases}$ 
\end{center}and the dimension $k$ is provided as follows:
\begin{itemize}
\item[(1)]If $q=3$ and $2\leq\delta\leq \frac{q^{h+1}+5}{4}$, then 
\begin{itemize}
\item[1.1)]If $m\equiv2\pmod4$, then $$k=n-m\big\lceil \frac{(2\delta-3)(q-1)}{2q}\big\rceil.$$
\item[1.2)]If $m\equiv0\pmod4$, then
$$k=\begin{cases}
n-m\big\lceil \frac{(2\delta-3)(q-1)}{2q}\big\rceil,&\ \text{if}\ 2\leq \delta\leq \frac{q^h+3}{4};\\
n-m\big\lceil \frac{(2\delta-3)(q-1)}{2q}\big\rceil+h,&\ \text{if}\ \frac{q^h+3}{4}+1\leq \delta\leq\frac{q^{h+1}+5}{4}.
\end{cases}$$	
\end{itemize}
			
\item[(2)]If $q>3$ and $m\equiv2\pmod4$, then 
\begin{itemize}
\item[2.1)]If $2\leq\delta\leq\frac{3(q^h+1)}{4}$, then $$k=n-m\big\lceil \frac{(2\delta-3)(q-1)}{2q}\big\rceil.$$
\item[2.2)]If $\frac{3(q^h+1)}{4}+1\leq \delta\leq\frac{q^{h+1}+5}{4}$, let $\tau\in\Z$ with $1\leq \tau\leq\frac{q-3}{4}$. Then
$$k=\begin{cases}
n-m\big(\big\lceil \frac{(2\delta-3)(q-1)}{2q}\big\rceil-(2\tau^2-\tau)\big),&\ \text{if}\ \tau (q^h+1)-\frac{q^h-3}{4}\leq \delta\leq \tau q^h+1;\\
n-m\big(\big\lceil \frac{(2\delta-3)(q-1)}{2q}\big\rceil-2\tau ^2\big),&\ \text{if}\ \tau (q^h+1)+1\leq \delta\leq \tau q^h+\frac{q^h+5}{4};\\
n-m\big(\big\lceil \frac{(2\delta-3)(q-1)}{2q}\big\rceil-(2\tau ^2+\tau )\big),&\ \text{if}\ \tau (q^h+1)+\frac{q^h+5}{4}\leq \delta\leq \tau q^h+\frac{q^h+3}{2};\\
n-m\big(\big\lceil \frac{(2\delta-3)(q-1)}{2q}\big\rceil-2(\tau ^2+\tau )\big),&\ \text{if}\ \tau (q^h+1)+\frac{q^h+3}{2}\leq \delta\leq \tau q^h+\frac{3(q^h+1)}{4}\ \text{and $\tau\neq \frac{q-3}{4}$};\\
n-m\big(\big\lceil \frac{(2\delta-3)(q-1)}{2q}\big\rceil-\frac{(q-3)(q+1)}{8}\big),&\ \text{if}\ \frac{(q-1)q^h}{4}+\frac{q+3}{4}\leq \delta\leq \frac{q^{h+1}+5}{4}.
\end{cases}$$	
\end{itemize}
\end{itemize}
\begin{proof}
Similar to Theorem \ref{t1}, we achieve the results using Lemmas \ref{Le3} and \ref{Le7}.
\end{proof}
\end{theorem}
\begin{example}
We give the following distance-optimal codes according to the Database of known best codes in (\cite{RefJ11}).
\\$\bullet$ Let $(q,m)=(3,4)$, then the codes  $\C_{(n,q,4)}$ and $\C_{(n,q,8)}$ has parameters $[20,14,4]$ and $[20,2,15]$, respectively.
\\$\bullet$ Let $(q,m)=(3,6)$, then the code $\C_{(n,q,5)}$ has parameters $[182,164,6]$. 
\end{example}
Next we will investigate the dimension of the code $\C_{(n,q,\delta)}$ with $2\leq \delta\leq \frac{3q^{\frac{m+1}{2}}-2q^{\frac{m-1}{2}}+3}{4}$, where $m=2h+1\geq 5$.
\begin{lemma}\label{le8}
Let $m=2h+1\geq 5$, $n=\frac{q^m-1}{2}$ and $a,b,c,d\in \Z$, define $f(a,b,c,d)=aq^{h+1}+bq^{h}+cq+d$ and $g(a,b,c,d)=aq^{h+1}+bq^{h}+\frac{q-1}{2}\sum_{l=2}^{h-1}q^l+cq+d$. We assume that
$$\begin{aligned}
A_1=&\{f(a,0,c,d):0\leq c<a\leq \frac{q-1}{2}, \ 1\leq d\leq q-1\},\\
A_2=&\{g(a,\frac{q-1}{2},c,d):1\leq c-\frac{q-1}{2}\leq a< \frac{q-1}{2}, \ 1\leq d\leq q-1\},\\
A_3=&\{g(a,\frac{q-1}{2},\frac{q-1}{2},d):0\leq a< \frac{q-1}{2} , \ \frac{q-1}{2}< d\leq q-1\},\\
A_4=&\{f(a,b,0,d):1\leq d\leq a<\frac{q-1}{2}, \ 1\leq b\leq q-1\},\\
A_5=&\{f(\frac{q-1}{2},b,0,d):1\leq d,b\leq\frac{q-1}{2} \},\\
A_6=&\{g(a,b,\frac{q-1}{2},d):1\leq d-\frac{q-1}{2}\leq a\leq \frac{q-1}{2}\ \text{and}\ 0\leq b<\frac{q-1}{2},\\
&or\ 0\leq d-\frac{q-1}{2}-1\leq a<\frac{q-1}{2}\ \text{and}\ \frac{q-1}{2}< b\leq q-1\}.
\end{aligned}$$
If $1\leq i\leq \frac{q^\frac{m+3}{2}-1}{2}$ satisfying $i\not\equiv0\pmod q$ and $i\notin \cup_{i=1}^{6}A_{i}$, then $i\in \MinRep_{n}$ and $|C_{i}^{n}|=m$.
		
\begin{proof}We first discuss the value of $|C_{i}^{n}|$, where $1\leq i\leq \frac{q^\frac{m+3}{2}-1}{2}$. For $m=5$ or 7, we have $|C_{i}^{n}|=1$ or $m$. Note that $i<iq<n$, then $|C_{i}^{n}|\neq1$, i.e., $|C_{i}^{n}|=m$. For $m\geq9$, then $ \frac{m}{3}\geq |C_{i}^{n}|$ by $|C_{i}^{n}|\mid m$. Note that $i<iq^k<n$ for any $k\in [1,\frac{m}{3}]$, then $|C_{i}^{n}|=m$.
			
To investigate all coset leaders $i$ such that $1\leq i\leq \frac{q^\frac{m+3}{2}-1}{2}$ with $i\not\equiv0\pmod q$, we need to solve the equation \begin{equation}\label{eq3}
iq^k\equiv j\pmod n,
\end{equation} where $1\leq j<i\leq \frac{q^\frac{m+3}{2}-1}{2}$, $1\leq k\leq 2h$ and $q\nmid i,j$. We denote the $q-$expansions of $i,j$ as  $i=\sum_{l=0}^{h+1}i_{l}q^l$, $j=\sum_{l=0}^{h+1}j_{l}q^l$, where $1\leq i_{0},j_0\leq q-1$, $0\leq i_{h+1},j_{h+1}\leq \frac{q-1}{2}$ and $0\leq i_l,j_l\leq q-1$ for any $l\in [1,h]$.
\\$\bf{Case\ 1.}$ For $1\leq k\leq h-1$, then $i<iq^k\leq \frac{q^{h+2}-1}{2}q^{h-1}<n$, i.e., the Equation (\ref{eq3}) has no solution.
\\$\bf{Case\ 2.}$ For $h+2\leq k\leq 2h$, it is easy to see that $$[iq^k]_{2n}=(i_{2h-k},i_{2h-k-1},\ldots,i_0,\underbrace{0,\ldots}_{h-1},i_{h+1},i_{h},\ldots,i_{2h-k+1}).$$
If $i_{2h-k}>\frac{q-1}{2}$, then
$$\frac{q^{h+2}-1}{2}\leq (0,\underbrace{\frac{q-1}{2},\ldots}_{m-1})<[iq^k]_{2n}-n=[iq^k]_{n} ,$$ i.e., the Equation (\ref{eq3}) has no solution.
\\If $\frac{q-1}{2}>i_{2h-k}$, then $\frac{q^{h+2}-1}{2}<(0,\ldots,0,i_0,\underbrace{0,\ldots}_{h-1},i_{h+1},\ldots,i_{2h-k+1})<[iq^k]_{2n}=[iq^k]_{n}$, i.e., the Equation (\ref{eq3}) has no solution.
\\For $i_{2h-k}=\frac{q-1}{2}$, we have the following.
\begin{itemize}
\item[1)]If $i_l= \frac{q-1}{2}$ for any $l\in [0,2h-k]$, then $[iq^k]_{n}=[iq^k]_{2n}>\frac{q^{h+2}-1}{2}$, i.e., the Equation (\ref{eq3}) has no solution.
\item[2)]If there exists $l_1\in\Z$ such that $i_{l_{1}}\neq\frac{q-1}{2}$ and $i_l=\frac{q-1}{2}$ for any $l\in [l_1+1,2h-k]$, then
\begin{center}
$[iq^k]_{n}=$
$\begin{cases}
[iq^k]_{2n}>\frac{q^{h+2}-1}{2},&\ \text{if}\  \frac{q-1}{2}>i_{l_{1}};\\
[iq^k]_{2n}-n>\frac{q^{h+2}-1}{2},&\ \text{if}\ i_{l_{1}}> \frac{q-1}{2},
\end{cases}$
\end{center}i.e., the Equation (\ref{eq3}) has no solution.
\end{itemize}
$\bf{Case\ 3.}$ For $k=h$, we have
\begin{equation}\label{eq4}
[iq^k]_{2n}=(i_{h},i_{h-1},\ldots,i_0,\underbrace{0,\ldots}_{h-1},i_{h+1}),
\end{equation} 
then we have the following.
\begin{itemize}
\item[1)] If $i_h>\frac{q-1}{2}$, then $[iq^k]_{n}=[iq^k]_{2n}-n>\frac{q^{h+2}-1}{2}$, i.e., the Equation (\ref{eq3}) has no solution.
\item[2)] If $\frac{q-1}{2}>i_h$, then $[iq^k]_{n}=[iq^k]_{2n}$. From (\ref{eq3}) and (\ref{eq4}), we have 
$$i=i_{h+1}q^{h+1}+i_1q+i_0,\ j=i_1q^{h+1}+i_0q^h+i_{h+1}.$$
Note that $i>j$, then Equation (\ref{eq3}) has a solution of $i=i_{h+1}q^{h+1}+i_1q+i_0$, where $1\leq i_0\leq q-1$ and $0\leq i_1<i_{h+1}\leq \frac{q-1}{2}$.
\item[3)] For $i_h=i_{h-1}=\cdots=i_0=\frac{q-1}{2}$ or there exists $l_1\in\Z$ such that $\frac{q-1}{2}>i_{l_{1}}$ and $i_l=\frac{q-1}{2}$ for any $l\in [l_1+1,h]$, then $[iq^k]_{n}=[iq^k]_{2n}>\frac{q^{h+2}-1}{2}$, i.e., the Equation (\ref{eq3}) has no solution.
\item[4)] If there exists $l_1\in\Z$ such that $i_{l_{1}}>\frac{q-1}{2}$ and $i_l=\frac{q-1}{2}$ for any $l\in [l_1+1,h]$, we have $\frac{q-1}{2}>i_{h+1}$ and $[iq^k]_{n}=[iq^k]_{2n}-n$.
\\If $l_{1}\geq 2$, then $$\frac{q^{h+2}-1}{2}<(0,\ldots,0,i_{1}+\frac{q-1}{2},i_0+\frac{q-1}{2},\underbrace{\frac{q-1}{2},\ldots}_{h-1},i_{h+1}+\frac{q-1}{2})<[iq^h]_n,$$ i.e., the Equation (\ref{eq3}) has no solution.
\\If $l_{1}= 1$, we have  $$i=(0,\ldots,0,i_{h+1},\underbrace{\frac{q-1}{2},\ldots}_{h-1},i_1,i_0),$$
\begin{center}
$\left\{\begin{aligned}
&[iq^h]_n=(0,\ldots,0,i_{1}-\frac{q-1}{2},i_0-\frac{q-1}{2}-1,\underbrace{\frac{q-1}{2},\ldots}_{h-1},i_{h+1}+\frac{q-1}{2}+1),&\ \text{if} \ i_0> \frac{q-1}{2};\\
&[iq^h]_n=(0,\ldots,0,i_{1}-\frac{q-1}{2}-1,i_0+\frac{q-1}{2},\underbrace{\frac{q-1}{2},\ldots}_{h-1},i_{h+1}+\frac{q-1}{2}+1),&\ \text{if} \ \frac{q-1}{2}\geq i_0.
\end{aligned} 
\right.$
\end{center}
Note that $i>j$, then Equation (\ref{eq3}) has a solution of $i=i_{h+1}q^{h+1}+\frac{q-1}{2}\sum_{l=2}^{h}q^l+i_1q+i_0$, where 
\begin{center}
$\begin{cases}
1\leq i_1-\frac{q-1}{2}\leq i_{h+1}<\frac{q-1}{2},&\ \text{if}\ \frac{q-1}{2}< i_0\leq q-1;\\
0\leq i_1-\frac{q-1}{2}-1< i_{h+1}<\frac{q-1}{2},&\ \text{if}\ 1\leq i_0\leq\frac{q-1}{2},
\end{cases}$ $\Rightarrow$ 
$\begin{cases}
1\leq i_1-\frac{q-1}{2}\leq i_{h+1}<\frac{q-1}{2},\\ 
and\ 1\leq i_0\leq q-1.
\end{cases}$
\end{center}
If $l_{1}= 0$, we have
\begin{center}
$\left\{\begin{aligned}	&i=(0,\ldots,0,i_{h+1},\frac{q-1}{2},\underbrace{\frac{q-1}{2},\ldots}_{h-1},i_0);\\
&[iq^h]_n=(0,\ldots,0,0,i_0-\frac{q-1}{2}-1,\underbrace{\frac{q-1}{2},\ldots}_{h-1},i_{h+1}+\frac{q-1}{2}+1).
\end{aligned} 
\right.$
\end{center}
Note that $i>j$, then Equation (\ref{eq3}) has a solution of $i=i_{h+1}q^{h+1}+\frac{q-1}{2}\sum_{l=1}^{h}q^l+i_0$, where $0\leq i_{h+1}<\frac{q-1}{2}$ and $\frac{q-1}{2}< i_0\leq q-1$.
\end{itemize}
$\bf{Case\ 4.}$ If $k=h+1$, then
$$[iq^k]_{2n}=(i_{h-1},i_{h-2},\ldots,i_0,\underbrace{0,\ldots}_{h-1},i_{h+1},i_{h}).$$Using the same approach as in Case 3, we can conclude that Equation (\ref{eq3}) has solutions of $i=i_{h+1}q^{h+1}+i_{h}q^h+i_0$, where $$\begin{cases}
1\leq i_0\leq i_{h+1}<\frac{q-1}{2},&\ \text{if} \ 1\leq i_h\leq q-1;\\
1\leq i_0\leq i_{h+1}=\frac{q-1}{2},&\ \text{if}\ 1\leq i_h\leq \frac{q-1}{2},
\end{cases}$$ and $i=i_{h+1}q^{h+1}+i_hq^h+\frac{q-1}{2}\sum_{l=1}^{h-1}q^l+i_0$, where
$$\begin{cases}
1\leq i_0-\frac{q-1}{2}\leq i_{h+1}\leq \frac{q-1}{2},\ &\ \text{if}\ 0\leq i_h<\frac{q-1}{2};\\ 
0\leq i_0-\frac{q-1}{2}-1\leq i_{h+1}<\frac{q-1}{2},\ &\ \text{if}\ \frac{q-1}{2}< i_h\leq q-1.
\end{cases}$$ This concludes our proof.
\end{proof}
\end{lemma}
\begin{theorem}\label{t3}
Let $m=2h+1\geq 5$, $n=\frac{q^m-1}{2}$ and $\epsilon=\lceil (\delta-1)(1-q^{-1})\rceil$. Then the cyclic code $\C^{'}_{(n,q,\delta,1)}$ has parameters $[n,k,d\geq \delta]$, where the dimension $k$ is provided as follows:
\begin{itemize}
\item [(1)] If $2\leq \delta\leq q^{h+1}+\frac{q-1}{2}q^h+1$, let $\tau\in\Z$ with $0\leq \tau\leq \frac{q-3}{2}$. Then 
\begin{center}
$k=$$\begin{cases}
n-m\epsilon,& \text{if}\ 2\leq \delta\leq\frac{q^{h+1}+1}{2};\\
n-m(\epsilon-\frac{q-1}{2}),& \text{if}\ \frac{q^{h+1}+q}{2}\leq \delta\leq\frac{q^{h+1}+1}{2}+q^h;\\
n-m(\epsilon-\frac{q-1}{2}-\tau),& \text{if}\ \frac{q^{h+1}+1}{2}+\tau q^h+1\leq \delta\leq\frac{q^{h+1}+1}{2}+(\tau+1)q^h,\tau\neq 0;\\
n-m(\epsilon-q+1),& \text{if}\ q^{h+1}-\frac{q^h-1}{2}+1\leq \delta\leq q^{h+1}+1;\\
n-m(\epsilon-2(q-1)),& \text{if}\ q^{h+1}+q\leq \delta\leq q^{h+1}+\frac{q^h+1}{2};\\
n-m(\epsilon-2(q-1)-(2\tau+1)),& \text{if}\ q^{h+1}+\frac{q^h+1}{2}+\tau q^h+1\leq \delta\leq q^{h+1}+(\tau+1)q^h+1;\\
n-m(\epsilon-2(q-1)-2\tau),& \text{if}\ q^{h+1}+\tau q^h+2\leq \delta\leq q^{h+1}+\frac{q^h+1}{2}+\tau q^h,\tau\neq 0.
\end{cases}$
\end{center}
\item [(2)] If $q=3$ and $q^{h+1}+\frac{q-1}{2}q^h+2\leq \delta\leq \frac{q^{h+2}+1}{2}$, then $$k=n-m(\epsilon-6).$$
\item [(3)] If $q\geq 5$ and $q^{h+1}+\frac{q-1}{2}q^h+2\leq \delta\leq2q^{h+1}-q^h+1$, let $\iota\in \Z$ with $\frac{q+1}{2}\leq \iota \leq q-2$. Then \begin{center}
$k=$$\begin{cases}
n-m(\epsilon-3(q-1)),& \text{if}\ q^{h+1}+\frac{q-1}{2}q^h+2\leq \delta\leq q^{h+1}+\frac{q^{h+1}+1}{2};\\
n-m(\epsilon-\frac{7(q-1)}{2}),& \text{if}\ q^{h+1}+\frac{q^{h+1}+q}{2}\leq \delta\leq q^{h+1}+\frac{(q+1)q^h}{2}+1;\\
n-m(\epsilon-3(q-1)-\iota),& \text{if}\ q^{h+1}+\iota q^h+2\leq \delta\leq q^{h+1}+(\iota+1)q^h+1.
\end{cases}$\end{center}
\end{itemize}
\begin{proof}
It is clear that the generator polynomal of $\C^{'}_{(n,q,\delta,1)}$ is $g(x)=\lcm \left(m_{\gamma^{1}}(x),m_{\gamma^{2}}(x),\ldots,m_{\gamma^{\delta-1}}(x)\right)$,  where $\gamma$ is a primitive $n$-th root of unity. Let $\Gamma^{'}=\lbrace i:1\leq i\leq \delta-1,\ \ i\not\equiv 0\pmod q\rbrace$, then $|\Gamma^{'}|=\lceil (\delta-1)(1-q^{-1})\rceil$, where $2\leq \delta\leq \frac{q^{h+2}+3}{2}$. For any $i\in \Gamma^{'}$, we have $|C_{i}^{n}|=m$ and $i\in\MinRep_{n}$ except that $i\in \cup_{l=1}^{6} A_{l}$ by Lemma \ref{le8}, where $A_{l}$ is given in Lemma \ref{le8}. Note that $A_{i}\cap A_{j}=\emptyset$ for $1\leq i\neq j\leq 6$, then
\begin{center}
$k=n-m|\Gamma^{'}|+m\sum_{l=1}^{6}|\Gamma^{'}\cap A_{l}|.$
\end{center}
We obtain the desired results using the same method as in Theorem \ref{t1}.
\end{proof}
\end{theorem}
\begin{theorem}\label{t4}
Let $q\equiv1\pmod4$, $m=2h+1\geq 5$ and $n=\frac{q^m-1}{4}$. Then the code $\C_{(n,q,\delta)}$ has parameters $[n,k,d]$, where \begin{center}
$\begin{cases}
d\geq\delta+1,& \ \text{if}\ \delta\equiv\frac{q+1}{2}\pmod q;\\
d\geq\delta,&\  \text{otherwise}
\end{cases}$ 
\end{center}and the dimension $k$ is provided as follows:
\begin{center}
$k=$$\begin{cases}
n-m\big\lceil \frac{(2\delta-3)(q-1)}{2q}\big\rceil,& \text{if}\ 2\leq \delta\leq\frac{q^{h+1}+3}{4};\\
n-m\big(\big\lceil \frac{(2\delta-3)(q-1)}{2q}\big\rceil-\frac{q-1}{4}\big),& \text{if}\ \frac{q^{h+1}+q+2}{4}\leq \delta\leq\frac{q^{h+1}+3}{4}+q^h;\\
n-m\big(\big\lceil \frac{(2\delta-3)(q-1)}{2q}\big\rceil-\frac{q-1}{4}-\tau\big),& \text{if}\ \frac{q^{h+1}+7}{4}+\tau q^h\leq \delta\leq\frac{q^{h+1}+3}{4}+(\tau+1)q^h,\tau \neq\frac{q-1}{4};\\
n-m\big(\big\lceil \frac{(2\delta-3)(q-1)}{2q}\big\rceil-\frac{q-1}{2}\big),& \text{if}\ \frac{2q^{h+1}-q^h+3}{4}+1\leq \delta\leq\frac{q^{h+1}+3}{2};\\
n-m\big(\big\lceil \frac{(2\delta-3)(q-1)}{2q}\big\rceil-(q-1)\big),& \text{if}\ \frac{q^{h+1}+q}{2}+1\leq \delta\leq\frac{q^{h+1}+q^h}{2}+1;\\
n-m\big(\big\lceil \frac{(2\delta-3)(q-1)}{2q}\big\rceil-(q+2\tau -2)\big),& \text{if}\ \frac{q^{h+1}-q^h}{2}+\tau q^h+2\leq \delta\leq\frac{2q^{h+1}-q^h+3}{4}+\tau q^h;\\
n-m\big(\big\lceil \frac{(2\delta-3)(q-1)}{2q}\big\rceil-(q+2\tau -1)\big),& \text{if}\ \frac{2q^{h+1}-q^h+7}{4}+\tau q^h\leq \delta\leq\frac{q^{h+1}+q^h}{2}+\tau q^h+1,\tau \neq\frac{q-1}{4},
\end{cases}$
\end{center} where $\tau\in\Z$ with $1\leq \tau \leq \frac{q-1}{4}$.
\begin{proof}
Similar to Theorem \ref{t1}, we achieve the results using Lemma \ref{le8}.
\end{proof}
\end{theorem}
\begin{example}
We give the following distance-optimal codes according to the Database of known best codes in (\cite{RefJ11}).
\\$\bullet$ Let $(q,m)=(3,5)$, then the cyclic BCH code $\C^{'}_{(n,q,5,1)}$ have parameters $[121,111,6]$.
\\$\bullet$ Let $(q,m)=(3,5)$, then the cyclic BCH code $\C^{'}_{(n,q,41,1)}$ have parameters $[121,16,61]$.
\end{example}
Next we will investigate the parameters of the code  $\C_{(n,q,\delta)}$ with one zero or two zeros.
\begin{theorem}\label{t5}
Let $q\equiv1\pmod4$ or $q\equiv3\pmod4$ and $2\mid m$, $m\geq 3$ and $n=\frac{q^m-1}{4}$. Then the code $\C_{(n,q,2)}$ has parameters $[n,n-m,d]$, where
\begin{center}
$\begin{cases}
d=3,\ \text{if}\ q=3\ or\ q=5\ \text{and}\ 2\nmid m;\\
d=2,\ \text{if}\ q>5\ or\ q=5\ \text{and}\ 2\mid m.
\end{cases}$
\end{center}
\begin{proof}
Note that the generator polynomal of $\C_{(n,q,2)}$ is $g(x)=M_{\beta}(x)$, where $\beta$ is a primitive $2n$-th root of unity. Since $|C_{1}^{2n}|=m$ for any $m\geq3$, then $\dim(\C_{(n,q,2)})=n-m$. From Lemmas \ref{Le1} and \ref{Le2}, we have $2\leq$d$(\C_{(n,q,2)})\leq 4$ for any $q$. If $d=4$, this implies $q^{m-1}\geq 1+\frac{(q^m-5)(q-1)}{4}$ by Lemma \ref{Le2}, which is impossible for $m\geq 3$. Hence, $2\leq $d$(\C_{(n,q,2)})\leq 3$.
			
If $q=3$ and $2\mid m$, then $3\in C_{1}^{2n}\Rightarrow\C_{(n,q,2)}=\C_{(n,q,3)}$. From Lemma \ref{Le1}, we have d$(\C_{(n,q,2)})\geq 3$, i.e., d$(\C_{(n,q,2)})=3$. 
			
If $q=5$, then $\gcd(n,2)=\gcd(m,2)$. For $2\mid m$, we have $\beta^\frac{n}{2}\in \F_5^*$ and $x^{\frac{n}{2}}-\beta^\frac{n}{2}\in \C_{(n,q,2)}$, then d$(\C_{(n,q,2)})=2$. For $2\nmid m$, suppose there exists a codeword $a_0+a_1x^s\in \C_{(n,q,2)}$, where $a_0,a_1 \in \F_5^*$ and $s\in[1,n-1]$. We have $x^s+a_1^{-1}a_0\in\C_{(n,q,2)}$, which implies $\beta^{4s}=1$. It follows that $2n\mid 4s$, then $n\mid s$ by $\gcd(n,2)=1$. This contradicts the fact that $s\in[1,n-1]$, then d$(\C_{(n,q,2)})\neq2$, then d$(\C_{(n,q,2)})=3$ for $2\nmid m$.
			
If $q>5$, it is clear that $\beta^{\frac{q^m-1}{q-1}}\in \F_q^{*}$, then there exist $a_0$, $a_1\in \F_q^*$ such that $a_0+a_1\beta^{\frac{q^m-1}{q-1}}=0$. Note that $\frac{q^m-1}{4}>\frac{q^m-1}{q-1}$, this implies that $a_0+a_1x^{\frac{q^m-1}{q-1}}\in \C_{(n,q,2)}$, then d$(\C_{(n,q,2)})=2$. 
\\Hence, we have concluded the proof.
\end{proof}
\end{theorem}
\begin{theorem}\label{t6}
Let $q\equiv1\pmod4$ or $q\equiv3\pmod4$ and $2\mid m$, $m\geq 3$ and $n=\frac{q^m-1}{4}$. Then the code $\C_{(n,q,3)}$ has parameters $[n,n-2m,d]$, where
$$\begin{cases}
d=3,&\ \text{if}\ q>5 \ \text{and}\ q\neq 9\ when\ 2\nmid m;\\
d=4,&\ \text{if}\ q=5\ \text{with}\ 2\mid m;\\
4\leq d\leq 5,&\ \text{if}\ q=5\ \text{with}\ 2\nmid m;\\
3\leq d\leq 4,&\ \text{if}\ q=9\ \text{with}\ 2\nmid m.
\end{cases}$$
\begin{proof}
Note that the generator polynomal of $\C_{(n,q,3)}$ is $g(x)=M_{\beta}(x)M_{\beta^3}(x)$, where $\beta$ is a primitive $2n$-th root of unity. Since $|C_{1}^{2n}|=|C_{3}^{2n}|=m$ for any $m\geq3$, then $\dim(\C_{(n,q,3)})=n-2m$. From Lemmas \ref{Le1} and \ref{Le2}, we have $3\leq $d$(\C_{(n,q,3)})\leq 5$ for $q= 5$ and $3\leq $d$(\C_{(n,q,3)})\leq 4$ for $q\geq 7$.
			
If $q=5$, then $5\in C_{1}^{2n}$, then $4\leq $d$(\C_{(n,q,3)})\leq5$ by Lemma \ref{Le1}. For $2\mid m$, we have $6\mid n$, then let $c(x)=1+\beta^{\frac{n}{2}}x^{\frac{n}{6}}+\beta^{\frac{n}{2}}x^{\frac{n}{2}}-x^{\frac{2n}{3}}$. It is clear that $c(\beta)=c(\beta^3)=0$ and $c(x)\in \F_{5}[x]$, then $c(x)\in\C_{(n,q,3)}$ and d$(c(x))=4$, then d$(\C_{(n,q,3)})=4$.
			
For $q>5$, we have the following.
\begin{itemize}
\item[1)] If $q>9 $ and $2\nmid m$, we assume that $s_1=\frac{4n}{q-1}$ and $s_2=\frac{8n}{q-1}$. It is clear that $s_1,s_2\in\Z$ with $0<s_{1},s_2<n$, suppose the system of equations is:
$$\begin{cases}
1+a_0\beta^{s_1}+a_1\beta^{s_2}=0;\\
1+a_0\beta^{3s_1}+a_1\beta^{3s_2}=0.
\end{cases}$$
Note that $\beta^{s_1},\beta^{s_2}\in \F_{q}^*$, then 
\begin{center}
$\begin{vmatrix}
\beta^{s_1} & \beta^{s_2}\\
\beta^{3s_1} & \beta^{3s_2}\\
\end{vmatrix}$=$\beta^{\frac{20n}{q-1}}(\beta^{\frac{8n}{q-1}}-1)\in \F_{q}^*.$
\end{center}
Then the equations have a unique solution $(a_0,a_1)\in (\F_q^*)^2$. It is clear that $1+a_0x^{s_1}+a_1x^{s_2}\in \C_{(n,q,3)}$, then d$(\C_{(n,q,3)})=3$.
\item[2)] If $q\geq 7$ and $2\mid m$, we can assume $s_1=\frac{2n}{q-1}$ and $s_2=\frac{4n}{q-1}$. Applying the same method as described above, there exist $b_0,b_1\in \F_q^*$ such that $1+b_0x^{s_1}+b_1x^{s_2}\in \C_{(n,q,3)}$, then d$(\C_{(n,q,3)})=3$.
\end{itemize}
This concludes our proof.
\end{proof}
\end{theorem}
\begin{remark}
 An $[n,k,d]$ linear code $\C$ is said optimal if it meets a bound for linear codes.
 \begin{itemize}
 \item Let $m\geq 3$, $q=3$ and $2\mid m$ or $q=5$ and $2\nmid m$, $n=\frac{q^m-1}{4}$. Then the code $\C_{(n,q,2)}$ has parameters $[n,n-m,3]$ and is distance-optimal with respect to the sphere-packing bound.
 \end{itemize}
\end{remark}
\begin{example}
We give the following distance-optimal codes according to the Database of known best codes in (\cite{RefJ11}).
\\$\bullet$ Let $(q,m)=(3,4)$, the code $\C_{(n,q,2)}=\C_{(n,q,3)}$ has parameters $[20,16,3]$.
\\$\bullet$ Let $(q,m)=(3,6)$, the code $\C_{(n,q,2)}=\C_{(n,q,3)}$ has parameters $[182,176,3]$.
\\$\bullet$ Let $(q,m)=(5,3)$, the code $\C_{(n,q,2)}$ has parameters $[31,28,3]$, and $\C_{(n,q,3)}$ has parameters $[31,25,4]$.
\end{example}
In the following, we will investigate the parameters of the code $\C_{(n,q,\delta)}$ with small dimension.
\begin{lemma}\label{le9}
Let $n=\frac{q^m-1}{2}$, then
\begin{itemize}
\item[(1)]If $q\equiv1\pmod4$, then the first three largest odd coset leaders modulo $n$ are:
\begin{center}
$\delta_{1}=\frac{q^m-q^{m-1}-q^{\lfloor\frac{m-1}{2}\rfloor}-1}{2},\ 
\delta_{2}=\frac{q^m-q^{m-1}-q^{\lfloor\frac{m+1}{2}\rfloor}-1}{2},\ 
\delta_{3}=\frac{q^m-q^{m-1}-q^{\lfloor\frac{m+3}{2}\rfloor}-1}{2}\  (m\geq 6).$
\end{center}
Moreover, $$\begin{cases}
|C_{\delta_{1}}^{n}|=\frac{m}{2},\ |C_{\delta_{2}}^{n}|=|C_{\delta_{3}}^{n}|=m,&\ \text{if}\  2\mid m;\\
|C_{\delta_{1}}^{n}|=|C_{\delta_{2}}^{n}|=|C_{\delta_{3}}^{n}|=m,&\ \text{if}\  2\nmid m.
\end{cases}$$	
\item[(2)]If $q\equiv3\pmod4$ and $2\mid m$, then the first two largest odd coset leaders modulo $n$ are:\begin{center}
$\delta_{1}=$$\begin{cases}
\frac{q^{m}-q^{m-1}-q^{\frac{m}{2}}-1}{2},&\ \text{if} \ m\equiv2 \pmod4;\\
\frac{q^{m}-q^{m-1}-q^{\frac{m-2}{2}}-1}{2},&\ \text{if} \ m\equiv0 \pmod4.
\end{cases}$\\
$\delta_{2}=$
$\begin{cases}
\frac{q^{m}-q^{m-1}-q^{\frac{m+4}{2}}-1}{2},&\ \text{if} \ m\equiv2 \pmod4\ with \ m\geq 10;\\
\frac{q^{m}-q^{m-1}-q^{\frac{m+2}{2}}-1}{2},&\ \text{if} \ m\equiv0 \pmod4\ with \ m\geq 6.
\end{cases}$
\end{center}
Moreover, 
$$\begin{cases}
|C_{\delta_{1}}^{n}|=|C_{\delta_{2}}^{n}|=m,&\ \text{if}\  m\equiv2\pmod 4;\\
|C_{\delta_{1}}^{n}|=\frac{m}{2},\ |C_{\delta_{2}}^{n}|=m,&\ \text{if}\  m\equiv0\pmod 4.
\end{cases}$$	
\end{itemize}
\begin{proof}
We only give the proof of $q\equiv1\pmod4$, since other cases are similar. It is easy to get that $\delta_{i}^{'}=\frac{q^m-q^{m-1}}{2}-\frac{q^{\lfloor\frac{m-3}{2}+i\rfloor}+1}{2}$ is the $i-$th largest coset leader by Lemma \ref{Le4}. Since $q\equiv1\pmod4$, we deduce $q^m\equiv1\pmod4$ for any $m$, then $\delta_{i}^{'}$ is odd for any $1\leq i\leq \lfloor\frac{m+6}{4}\rfloor$. 
\\This concludes our proof.
\end{proof}
\end{lemma}
\begin{theorem}\label{t7}
Let $q\equiv1\pmod4$, $n=\frac{q^m-1}{4}$ and $\delta_1,\delta_2,\delta_3$ be given in Lemma \ref{le9}. Then the code $\C_{(n,q,\delta)}$ has parameters $[n,m(i-1)+\kappa,d\geq \frac{\delta_{i}+1}{2}]$ with  $\frac{\delta_{i+1}+3}{2}\leq \delta\leq \frac{\delta_{i}+1}{2}$ $(i=1,2)$, where \begin{center}
$\kappa=$$\begin{cases}
\frac{m}{2},&\ \text{if}\ 2\mid m;\\
m,&\ \text{if}\ 2\nmid m .
\end{cases}$	
\end{center}
\begin{proof}
If $\frac{\delta_{2}+3}{2}\leq \delta\leq \frac{\delta_{1}+1}{2},$ then the check polynomial of $\C_{(n,q,\delta)}$ is $M_{\beta^{\delta_{1}}}(x)$, where $\beta$ is a primitive $2n$-th root of unity. By utilizing Lemmas \ref{Le1} and \ref{le9}, we can arrive at the expected conclusion. The proof of $\frac{\delta_{3}+3}{2}\leq \delta\leq \frac{\delta_{2}+1}{2}$ is similar, hence is omitted here.
\end{proof}
\end{theorem}
\begin{theorem}\label{t8}
Let $q\equiv3\pmod4$, $ m\equiv2 \pmod4\ with \ m\geq 10$ and $n=\frac{q^m-1}{4}$. If $\frac{\delta_{2}+3}{2}\leq \delta\leq \frac{\delta_{1}+1}{2}$, $\delta_{1},\delta_2$ are given in Lemma \ref{le9}, then 
\begin{itemize}
\item[(1)]If $q>3$, then the code $\C_{(n,q,\delta)}$ has parameters $[n,m,d\geq \frac{\delta_{1}+1}{2}]$.
\item[(2)]If $q=3$, then $\dim(\C_{(n,3,\delta)})=m$ and $\C_{(n,3,\delta)}$ is a two-weight code. Moreover, the weight distribution of $\C_{(n,3,\delta)}$ is presented in Table 1.
\begin{table}[h]
\centering
\footnotesize
\renewcommand{\arraystretch}{1.5}
\setlength{\tabcolsep}{6pt}
\caption{Weight distribution of $\C_{(n,3,\delta)}$}
\label{tab:my_label}
\scalebox{1.0}{
\begin{tabular}{lll}
\toprule
\textbf{Weight} & \textbf{Frequency} \\ \midrule
0 & 1 \\
$\frac{3^{m-1}-3^\frac{m-2}{2}}{2}$ & $\frac{3^m-1}{2}$ \\
$\frac{3^{m-1}+3^\frac{m-2}{2}}{2}$ & $\frac{3^m-1}{2}$ \\
\bottomrule
\end{tabular}
}
\end{table}
\end{itemize}
\begin{proof}
If $\frac{\delta_{2}+3}{2}\leq \delta\leq \frac{\delta_{1}+1}{2}$, we can get $\dim(\C_{(n,q,\delta)})=m$ and d$(\C_{(n,q,\delta)})\geq \frac{\delta_{1}+1}{2}$ by Lemmas \ref{Le1} and \ref{le9}. Note that the code $\C_{(n,3,\delta)}$ has one nonzero $\beta^{\delta_{1}}$, where $\beta$ is a primitive $2n$-th root of unity in $\F_{3^m}$. Let $\alpha$ be a primitive element of $\F_{3^m}$, then $\beta=\alpha^{2}$. Note that $-2\delta_{1}\equiv 3^{m-1}+3^{\frac{m}{2}} \pmod {3^{m}-1}$, then the trace expression of $\C_{(n,3,\delta)}$ is
$$\begin{aligned}
\C_{(n,3,\delta)}&=\{(Tr_{3^{m}/3}(a\beta^{-\delta_{1}i}))_{i=0}^{n-1}: a\in \F_{3^m}\}\\
&=\{(Tr_{3^{m}/3}(a\alpha^{(3^{m-1}+3^{\frac{m}{2}})i}))_{i=0}^{n-1}: a\in \F_{3^m}\}\end{aligned} $$  
Note that $Tr_{3^{m}/3}(a\alpha^{(3^{m-1}+3^{\frac{m}{2}})i})=Tr_{3^{m}/3}(a^3\alpha^{(3^{\frac{m+2}{2}}+1)i})$ and $\gcd(3^{m}-1,3^{\frac{m+2}{2}}+1)=2$, then we give the following code which has the same weight distribution with $\C_{(n,3,\delta)}$ $$\{c(a)=(Tr_{3^{m}/3}(a\alpha^{2i}))_{i=0}^{n-1}: a\in \F_{3^m}\}.$$
			
Clearly, $c(0)$ is the zero codeword. If $a\neq0$, we have 
$$\begin{aligned}\label{eq1}
wt(c(a))&=n-|\{i:Tr_{3^{m}/3}(a\alpha^{2i})=0,0\leq i\leq n-1\}|\\
&=n-\frac{1}{3}\sum_{i=0}^{n-1}\sum_{x\in \F_{3}}\zeta_{p}^{xTr_{3^{m}/3}(a\alpha^{2i})}\\
&=\frac{2n}{3}-\frac{1}{3}\sum_{i=0}^{n-1}\sum_{x\in \F_{3}^*}\zeta_{p}^{xTr_{3^{m}/3}(a\alpha^{2i})}.
\end{aligned}$$
Note that $q=3$, then 
$$\begin{aligned}
\sum_{i=0}^{n-1}\sum_{x\in \F_{3}^*}\zeta_{p}^{xTr_{3^{m}/3}(a\alpha^{2i})}&=\sum_{i=0}^{n-1}\zeta_{p}^{Tr_{3^{m}/3}(a\alpha^{2i})}+\sum_{i=0}^{n-1}\zeta_{p}^{-Tr_{3^{m}/3}(a\alpha^{2i})}\\
&=\sum_{i=0}^{n-1}\zeta_{p}^{Tr_{3^{m}/3}(a\alpha^{2i})}+\sum_{i=0}^{n-1}\zeta_{p}^{Tr_{3^{m}/3}(a\alpha^{2(n+i)})}\\
&=\sum_{i=0}^{2n-1}\zeta_{p}^{Tr_{3^{m}/3}(a\alpha^{2i})}.
\end{aligned}$$
Note that $\sum_{i=0}^{2n-1}\zeta_{p}^{Tr_{3^{m}/3}(a\alpha^{2i})}=\sum_{i=2n}^{4n-1}\zeta_{p}^{Tr_{3^{m}/3}(a\alpha^{2i})}$, then
$$\begin{aligned}
wt(c(a))&=\frac{2n}{3}-\frac{1}{3}\sum_{i=0}^{n-1}\sum_{x\in \F_{3}^*}\zeta_{p}^{xTr_{3^{m}/3}(a\alpha^{2i})}\\
&=\frac{2n}{3}-\frac{1}{3}\sum_{i=0}^{2n-1}\zeta_{p}^{Tr_{3^{m}/3}(a\alpha^{2i})}\\
&=\frac{2n}{3}-\frac{1}{6}\sum_{i=0}^{4n-1}\zeta_{p}^{Tr_{3^{m}/3}(a\alpha^{2i})}\\
&=\frac{2n}{3}-\frac{1}{6}\sum_{y\in \F_{3^{m}}^{*}}\zeta_{p}^{Tr_{3^{m}/3}(ay^{2})}\\
&=\frac{3^m-1}{6}-\frac{1}{6}(\eta(a)G(\eta)-1),
\end{aligned}$$
where $\eta$ is the quadratic character of $\F_{3^m}$ and $G(\eta)$ is the quadratic Gauss sum. For all $a\in  \F_{3^{m}}$, we have $\eta(a)=1$ for $\frac{3^m-1}{2}$ times, and $\eta(a)=-1$ for $\frac{3^m-1}{2}$ times.
Using the value of $G(\eta)=(-1)^{m-1}(\sqrt{-1})^mq^{\frac{m}{2}}$, we obtain the weight distribution as Table $1$. 
\\Hence, we have concluded the proof.
\end{proof}
\end{theorem}
\begin{theorem}\label{t9}
Let $q\equiv3\pmod4$, $ m\equiv0 \pmod4\ with \ m\geq 6$ and $n=\frac{q^m-1}{4}$. If $\frac{\delta_{2}+3}{2}\leq \delta\leq \frac{\delta_{1}+1}{2}$, then the code $\C_{(n,q,\delta)}$ is an  $[n,\frac{m}{2},\frac{(q-1)(q^{m-1}+q^{\frac{m-2}{2}})}{4}]$ one-weight code, $\delta_1,\delta_2$ are given in Lemma \ref{le9}.
\begin{proof}
If $\frac{\delta_{2}+3}{2}\leq \delta\leq \frac{\delta_{1}+1}{2}$, we can get $\dim(\C_{(n,q,\delta)})=\frac{m}{2}$ by Lemma \ref{le9}. Note that the code $\C_{(n,q,\delta)}$ has only one nonzeroes $\beta^{\delta_{1}}$, where $\beta$ is a primitive $2n$-th root of unity in $\F_{q^m}$. Denote $\alpha$ is a primitive element of $\F_{q^m}$, then $\beta=\alpha^{2}$. Let $h=\frac{m}{2}$, then $2\mid h$. Note that $-2\delta_{1}\equiv q^{m-1}+q^{\frac{m-2}{2}} \pmod {q^{m}-1}$ and $|C_{\delta_1}^{2n}|=\frac{m}{2}$, then the trace expression of $\C_{(n,q,\delta)}$ is
$$\begin{aligned}
\C_{(n,q,\delta)}&=\{(Tr_{q^{h}/q}(a\beta^{-\delta_{1}i}))_{i=0}^{n-1}: a\in \F_{q^h}\}\\
&=\{(Tr_{q^{h}/q}(a\alpha^{(q^{m-1}+q^{\frac{m-2}{2}})i}))_{i=0}^{n-1}: a\in \F_{q^h}\}.\end{aligned} $$ 
Note that $Tr_{q^{h}/q}(a\alpha^{(q^{m-1}+q^{\frac{m-2}{2}})i})=Tr_{q^{h}/q}(a^q\alpha^{(q^{\frac{m}{2}}+1)i})$, then we give the following code which has the same weight distribution with $\C_{(n,q,\delta)}$
$$\{c(a)=(Tr_{q^{h}/q}(a\alpha^{(q^{h}+1)i}))_{i=0}^{n-1}: a\in \F_{q^h}\}.$$
Note that $4\mid q^h-1$, then $\frac{q^{m}-1}{4}=\frac{q^h-1}{4}(q^h-1)+\frac{q^h-1}{2}$. Let $n^{'}=q^{h}-1$, we have $$c(a)=\underbrace{c_{1}(a)||\cdots||c_1(a)}_{\frac{q^h-1}{4}}||c_{2}(a),$$
where $||$ denotes the concatenation of vectors, $\C_{1}=\{c_1(a)=(Tr_{q^{h}/q}(a\alpha^{(q^{h}+1)i}))_{i=0}^{n^{'}-1}: a\in \F_{q^h}\}$ and $\C_{2}=\{c_2(a)=(Tr_{q^{h}/q}(a\alpha^{(q^{h}+1)i}))_{i=0}^{\frac{n^{'}}{2}-1}: a\in \F_{q^h}\}$.
Denote $\gamma=\alpha^{(q^{h}+1)}$, then $\gamma$ is a primitive element of $\F_{q^{h}}$. Then 
$$\begin{aligned}
wt(c_{1}(a))&=n^{'}-|\{i:Tr_{q^{h}/q}(a\gamma ^i)=0,0\leq i\leq n^{'}-1\}|\\
&=n^{'}-|\{x\in \F_{q^{h}}^{*}:Tr_{q^{h}/q}(ax)=0\}|.
\end{aligned}$$
Hence, $\C_{1}$ is a one-weight code over $\F_{q}$ with $wt(c_{1}(a))=q^{h}-q^{h-1}$. For $\C_{2}$, we have $$
\begin{aligned}
wt(c_{2}(a))&=\frac{n^{'}}{2}-|\{i:Tr_{q^{h}/q}(a\gamma ^i)=0,0\leq i\leq \frac{n^{'}}{2}-1\}|\\
&=\frac{n^{'}}{2}-\frac{1}{q}\sum_{i=0}^{\frac{n^{'}}{2}-1}\sum_{x\in \F_{q}}\zeta_{p}^{xTr_{q^{h}/q}(a\gamma^{i})}\\
&=\frac{n^{'}}{2}-\frac{1}{2q}\sum_{i=0}^{n^{'}-1}\sum_{x\in \F_{q}}\zeta_{p}^{xTr_{q^{h}/q}(a\gamma^{i})}\\
&=\frac{n^{'}}{2}-\frac{1}{2q}\sum_{x\in \F_{q}}\sum_{y\in \F_{q^h}^{*}}\zeta_{p}^{Tr_{q^{h}/q}(xay)}\\
&=\frac{n^{'}}{2}-\frac{n^{'}}{2q}+\frac{q-1}{2q}-\frac{1}{2q}\sum_{x\in \F_{q}^{*}}\sum_{y\in \F_{q^h}}\zeta_{p}^{Tr_{q^{h}/q}(xay)}\\
&=\frac{(q-1)q^{h-1}}{2}.
\end{aligned}$$
In summary, based on the above discussions, we have $$wt(c(a))=\frac{q^{h}-1}{4}wt(c_{1}(a))+wt(c_{2}(a))=\frac{(q^{2h-1}+q^{h-1})(q-1)}{4}.$$ This concludes our proof.
\end{proof}	
\end{theorem}
\begin{theorem}\label{t10}
Let $q\equiv1\pmod4$, $m=2h\geq 2$ and $n=\frac{q^m-1}{4}$. If $\frac{\delta_{2}+3}{2}\leq \delta\leq \frac{\delta_{1}+1}{2}$, then the code $\C_{(n,q,\delta)}$ is an  $[n,\frac{m}{2},\frac{(q-1)(q^{m-1}+q^{\frac{m-2}{2}})}{4}]$ one-weight code, $\delta_1,\delta_2$ are given in Lemma \ref{le9}.
\begin{proof}
Similar to Theorem \ref{t9}, we achieve the results using Lemma \ref{le9}.
\end{proof}
\end{theorem}
\section{The case of $n=\frac{q^m+1}{4}$}
In this section, we consistently assume $q\equiv3\pmod 4$ and $2\nmid m$ when $n=\frac{q^m+1}{4}$. We first investigate the parameters of the code $\C_{(n,q,\delta)}$ with large dimension.
\begin{lemma}\label{le10}
Let $ q\equiv3\pmod 4$, $m=2h+1\geq 5$ and $n=\frac{q^m+1}{2}$.
\begin{itemize} 
\item If $2\mid h$, define$$\begin{aligned}
&B_1=\{\frac{q^{h+1}-1}{2}+2u:-\frac{q-3}{4}\leq u\leq \frac{q-3}{4}\},\\
&B_2=\begin{cases}
	\{\frac{q^{h+1}+1}{2}+\frac{q-1}{2}q^h\},&\ \text{if}\ q=3;\\
	\{\frac{q^{h+1}+1}{2}+(2v-1)q^h,\frac{q^{h+1}-1}{2}+2vq^h,\frac{q^{h+1}+1}{2}+\frac{q-1}{2}q^h:1\leq v\leq\frac{q-3}{4}\},&\ \text{if}\ q\neq3.
\end{cases}\end{aligned}$$
\item If $2\nmid h$, define
$$\begin{aligned}
&B_1=\{\frac{q^{h+1}+1}{2}+2u:-\frac{q-3}{4}\leq u\leq \frac{q-3}{4}\},\\
&B_2=\begin{cases}
	\{\frac{q^{h+1}-1}{2}+\frac{q-1}{2}q^h\},&\ \text{if}\ q=3;\\
	\{\frac{q^{h+1}-1}{2}+(2v-1)q^h,\frac{q^{h+1}+1}{2}+2vq^h,\frac{q^{h+1}-1}{2}+\frac{q-1}{2}q^h:1\leq v\leq\frac{q-3}{4}\},&\ \text{if}\ q\neq3.
\end{cases}\end{aligned}$$
\end{itemize}If $1\leq i\leq q^{h+1}-q$ is odd satifying $i\not\equiv0\pmod{q}$ and $i\notin B_1\cup B_2$, then $i\in \MinRep_{n}$.
\begin{proof}
Similar to Lemma \ref{Le7}, we achieve the results using Lemma \ref{Le5}.
\end{proof}
\end{lemma}
\begin{theorem}\label{t11}
Let $q\equiv3\pmod 4$, $m=2h+1\geq 5$ and $n=\frac{q^m+1}{2}$. Then the code $\C_{(n,q,\delta)}$ has parameters $[n,k,d]$, where \begin{center}
$\begin{cases}
d\geq 2\delta+1,& \ \text{if}\ \delta\equiv\frac{q+1}{2}\pmod q;\\
d\geq 2\delta-1,&\  \text{otherwise}
\end{cases}$ 
\end{center}and the dimension $k$ is provided as follows:
\begin{itemize}
\item[(1)]If $2\mid h$, let $\tau=0$ for $q=3$ and $\tau\in [1,\frac{q-3}{4}]$ for $q\neq 3$. Then 
$$k=\begin{cases}
n-2m\big\lceil \frac{(2\delta-3)(q-1)}{2q}\big\rceil;&\text{if}\ 2\leq \delta\leq \frac{q^{h+1}-q}{4}+1;\\
n-2m\big(\big\lceil \frac{(2\delta-3)(q-1)}{2q}\big\rceil-\frac{q-1}{2}\big), &\text{if}\ \frac{q^{h+1}+q+2}{4}\leq \delta\leq\frac{q^{h+1}+2q^h+3}{4};\\
n-2m\big(\big\lceil \frac{(2\delta-3)(q-1)}{2q}\big\rceil-\frac{q-1}{2}-(2\tau-1)\big),&\text{if}\ \frac{q^{h+1}-2q^h+3}{4}+\tau q^h+1\leq \delta\leq \frac{q^{h+1}+1}{4}+\tau q^h;\\
n-2m\big(\big\lceil \frac{(2\delta-3)(q-1)}{2q}\big\rceil-\frac{q-1}{2}-2\tau\big), &\text{if}\ \frac{q^{h+1}+5}{4}+\tau q^h\leq \delta\leq \frac{q^{h+1}+2q^h+3}{4}+\tau q^h;\\
n-2m\big(\big\lceil \frac{(2\delta-3)(q-1)}{2q}\big\rceil-(q-1)\big), &\text{if}\ \frac{2q^{h+1}-q^h+3}{4}+1\leq \delta\leq \frac{q^{h+1}-q}{2}+1.
\end{cases}$$
\item[(2)]If $2\nmid h$, let $\tau=0$ for $q=3$ and $\tau\in [1,\frac{q-3}{4}]$ for $q\neq 3$. Then 
$$k=\begin{cases}
n-2m\big\lceil \frac{(2\delta-3)(q-1)}{2q}\big\rceil,&\text{if}\ 2\leq \delta\leq \frac{q^{h+1}-q+2}{4}+1;\\
n-2m\big(\big\lceil \frac{(2\delta-3)(q-1)}{2q}\big\rceil-\frac{q-1}{2}\big), &\text{if}\ \frac{q^{h+1}+q}{4}+1\leq \delta\leq\frac{q^{h+1}+2q^h+1}{4};\\
n-2m\big(\big\lceil \frac{(2\delta-3)(q-1)}{2q}\big\rceil-\frac{q-1}{2}-(2\tau-1)\big),&\text{if}\ \frac{q^{h+1}-2q^h+1}{4}+\tau q^h+1\leq \delta\leq \frac{q^{h+1}+3}{4}+\tau q^h;\\
n-2m\big(\big\lceil \frac{(2\delta-3)(q-1)}{2q}\big\rceil-\frac{q-1}{2}-2\tau\big), &\text{if}\ \frac{q^{h+1}-1}{4}+\tau q^h+2\leq \delta\leq \frac{q^{h+1}+2q^h+1}{4}+\tau q^h;\\
n-2m\big(\big\lceil \frac{(2\delta-3)(q-1)}{2q}\big\rceil-(q-1)\big), &\text{if}\ \frac{2q^{h+1}-q^h+1}{4}+1\leq \delta\leq \frac{q^{h+1}-q}{2}+1.
\end{cases}$$	
\end{itemize}
\begin{proof}
We only prove the case of $2\mid h$, since the case of $2\nmid h$ is similar. Clearly, the generator polynomal of $\C_{(n,q,\delta)}$ is $g(x)=\lcm(M_{\beta^{1}}(x),M_{\beta^{3}}(x),\ldots,M_{\beta^{1+2(\delta-2)}}(x))$,  where $\beta$ is a primitive $2n$-th root of unity. Let $\Gamma=\lbrace 1+2i:0\leq i\leq \delta-2,\ \ 1+2i\not\equiv0\pmod q\rbrace$ for any $2\leq \delta\leq \frac{q^{h+1}-q}{2}+1$, then $|\Gamma|=\big\lceil\frac{(2\delta-3)(q-1)}{2q}\big\rceil.$ For any $1+2i\in \Gamma$, we have $1+2i\in \MinRep_{2n}$ except that $1+2i\in B_{1}\cup B_{2}$ and $|C_{1+2i}^{2n}|=2m$ by Lemmas \ref{Le5} and \ref{le10}. Then  
\begin{equation}\label{eq6}
k=n-2m|\Gamma|+2m(|\Gamma\cap B_1|+|\Gamma\cap B_2|).
\end{equation}
			
If $2\leq \delta\leq\frac{q^{h+1}-q}{4}+1$, note that $\min\{B_1\cup B_2\}=\frac{q^{h+1}-q}{2}+1>\max\{\Gamma\}$, then $\Gamma\cap B_1=\Gamma\cap B_2=\emptyset$. It follows from (\ref{eq6}) that $$k=n-2m\big\lceil \frac{(2\delta-3)(q-1)}{2q}\big\rceil.$$
Using the same approach, the results can be established for other cases as well.
\\Thus, we conclude the proof.
\end{proof}
\end{theorem}
\begin{theorem}\label{t12}
Let $q\equiv3\pmod4$, $m=2h+1\geq 3$ and $n=\frac{q^m+1}{4}$. Then the code $\C_{(n,q,2)}$ has parameters $[n,n-2m,d]$, where
$$\begin{cases}
d=3,&\ \text{if}\ q>7;\\
5\leq d\leq6,&\ \text{if}\ q=3;\\
3\leq d\leq4,&\ \text{if}\ q=7.
\end{cases}$$
\begin{proof}
Note that the generator polynomal of $\C_{(n,q,2)}$ is $g(x)=M_{\beta}(x)$, where $\beta$ is a primitive $2n$-th root of unity. Since $|C_{1}^{2n}|=2m$ for any $m\geq3$, then $\dim(\C_{(n,q,2)})=n-2m$. From Lemmas \ref{Le1} and \ref{Le2}, we have $3\leq $d$(\C_{(n,q,2)})\leq 6$ for $q= 3$ and $3\leq $d$(\C_{(n,q,2)})\leq 4$ for $q\geq 7$.
		
If $q=3$, we have $-3,-1,1,3\in C_1^{2n}$, then we have $5\leq $d$(\C_{(n,q,2)})\leq 6$ by Lemma \ref{Le1}.
			
If $q>7$, let $\nu=\beta^{\frac{q^m+1}{q+1}}$, then $\nu\in \F_{q^{2}}^{*}$ and $n>2\frac{q^m+1}{q+1}$. Then there exist $a_0,a_1,a_2\in \F_{q}^{*}$ such that $a_{0}+a_1\nu+a_2\nu^2=0$. It is clear that $\nu^2\neq \nu$ and $\nu^2\neq 1$, then $a_{0}+a_1x^{\frac{q^m+1}{q+1}}+a_2x^{2\frac{q^m+1}{q+1}} \in\C_{(n,q,2)}$, then d$(\C_{(n,q,2)})=3$.
\\This concludes our proof.
\end{proof}
\end{theorem}
\begin{example}
We give the following distance-optimal codes according to the Database of known best codes in (\cite{RefJ11}).
\\$\bullet$ Let $(q,m)=(3,5)$, the code $\C_{(n,q,2)}=\C_{(n,q,3)}$ has parameters $[61,51,5]$.
\\$\bullet$ Let $(q,m)=(7,3)$, the code $\C_{(n,q,2)}$ has parameters $[86,80,4]$.
\end{example}
In the following, we will investigate the parameters of  code $\C_{(n,q,\delta)}$ with small dimension.
\begin{lemma}\label{le11}
Let $a,u_k\in\Z$ with $0\leq a\leq \frac{q^m+1}{2}$, $n=q^m+1$ and $aq^k=u_k(q^m+1)+[aq^{k}]_{n}$. For any $1\leq k\leq m-1$, there is no $l\in\Z$ such that $$	u_{k}+\frac{[aq^{k}]_{n}-a}{q^{m}+1}<l<u_{k}+\frac{[aq^{k}]_{n}+a}{q^{m}+1},$$ then $a\in \MinRep_{n}$.
\begin{proof}
From Lemma \ref{Le6}, we have $a\in \MinRep_{n}$ and $0\leq a\leq  \frac{q^{m}+1}{2}$ which means that there are no integers $1\leq k\leq m-1$, $1\leq l\leq \frac{q^{k}-1}{2}$ and $-\frac{l(q^{m-k}-1)}{q^{k}+1}<h<\frac{l(q^{m-k}+1)}{q^{k}-1}$ such that 
\begin{equation}\label{equ1}
a=lq^{m-k}+h.
\end{equation}
If (\ref{equ1}) holds, we have
$\frac{l(q^{m}+1)}{q^{k}+1}<a<\frac{l(q^{m}+1)}{q^{k}-1}$ $\Rightarrow$ $\frac{a(q^{k}-1)}{q^{m}+1}<l<\frac{a(q^{k}+1)}{q^{m}+1}.$ Note that $aq^{k}=u_{k}(q^{m}+1)+[aq^{k}]_{n}$, then 
\begin{equation}\label{equ3}
u_{k}+\frac{[aq^{k}]_{n}-a}{q^{m}+1}<l<u_{k}+\frac{[aq^{k}]_{n}+a}{q^{m}+1}.
\end{equation}
Hence, if there is no $l\in\Z$ such that (\ref{equ3}) holds for any $1\leq k\leq m-1$, then $a\in \MinRep_{n}$.
\end{proof}
\end{lemma}
\begin{lemma}\label{le12}
Let $q\equiv3\pmod{8}$, $m=2h+1\geq 5$ and $n=\frac{q^m+1}{2}$. Then the first three largest odd coset leaders modulo $n$ are:
$$\delta_{1}=\frac{q^{m}+1}{4},\ \delta_{2}=\frac{q^{m}+1}{4}-\frac{q^{m-1}+q}{2},\ 
\delta_{3}=\frac{q^{m}-1}{4}-\frac{q^{m-1}+q^2-q}{2}.$$ 
Moreover, $|C_{\delta_{1}}^{n}|=1$ and $|C_{\delta_{2}}^{n}|=|C_{\delta_{3}}^{n}|=2m$.
\begin{proof}
Note that $q\equiv3\pmod{8}$ and $m=2h+1\geq 5$, then $q^m\equiv3\pmod8$ and $q^{m-1}\equiv1\pmod8$. Therefore $\delta_{1}^{'} $ and $\delta_{3}^{'}$ are odd, $\delta_{2}^{'}$ is even, where $\delta_{1}^{'}$, $\delta_{2}^{'}$, $\delta_{3}^{'}$ are given in Lemma \ref{Le5}. Hence, $\delta_1,\delta_2$ are the first two largest odd coset leaders. 
			
We claim that $\delta_{3} \in \MinRep_{n}$, i.e., $2\delta_{3}\in \MinRep_{2n}$ by Lemma \ref{Le6}. It is easy to get that
\begin{center}
$[2\delta_{3}q^{k}]_{2n}$=$\begin{cases}
\frac{q^m-1}{2}+q^{k+1}+q^{k-1}-q^{k+2}-q^k+1,& \text{if}\ 1\leq k\leq m-3;\\
\frac{q^m-1}{2}+q^{m-1}+q^{m-3}-q^{m-2}+2,& \text{if}\ k= m-2;\\
\frac{q^m-1}{2}+q^{m-2}-q^{m-1}+q,& \text{if}\ k=m-1.
\end{cases}$
\end{center}
On one hand, \begin{center}
$[2\delta_{3}q^{k}]_{2n}-2\delta_3$=$\begin{cases}
q^{m-1}+q^{k+1}+q^{k-1}-q^{k+2}-q^k+q^2-q+1>0,& \text{if}\ 1\leq k\leq m-3;\\
2q^{m-1}+q^{m-3}-q^{m-2}+q^2-q+2>0,& \text{if}\ k= m-2;\\
q^{m-2}+q^{2}>0,& \text{if}\ k=m-1.	
\end{cases}$
\end{center}
On the other hand, \begin{center}
$[2\delta_{3}q^{k}]_{2n}+2\delta_3$=$\begin{cases}
q^{m}+q^{k+1}+q^{k-1}-(q^{m-1}+q^{k+2}+q^k)-q^2+q<q^m+1 ,& \text{if}\ 1\leq k\leq m-3;\\
q^{m}+1-q^{m-2}-q^{m-3}-q^2+q<q^m+1,& \text{if}\ k= m-2;\\
q^m+q^{m-2}-2q^{m-1}-q^{2}+2q-1<q^m+1,& \text{if}\ k=m-1.
\end{cases}$
\end{center}
Hence, there is no $l\in\Z$ such that (\ref{equ3}) holds for any $k\in [1,m-1]$, i.e., $2\delta_3\in \MinRep_{2n}$ by Lemma \ref{le11}. Note that  $[2\delta_{3}q^{m}]_{2n}=\frac{q^m+3}{2}+q^{m-1}+q^2-q>2\delta_3$ and $[2\delta_{3}q^{k}]_{2n}-2\delta_3>0$ for any $k\in [1,m-1]$, then $|C_{2\delta_3}^{2n}|=2m$. Furthermore, $\delta_3$ is odd, then $\delta_3$ is odd coset leader modulo $n$ and $|C_{\delta_3}^{n}|=2m$.
			
Next, we will prove that there is no odd integer $a$ such that $a\in \MinRep_{n}$ for any  $\delta_3+1\leq a\leq\delta_2-1$. 
Suppose there is an odd integer $a$ such that $a\in \MinRep_{n}$ and $\delta_3+1\leq a\leq\delta_2-1$. From Lemma \ref{Le6}, we have $2a\in\MinRep_{2n}$ and $2\delta_3+2\leq 2a\leq2\delta_2-2$. Note that 
\begin{center}
$\left\{\begin{aligned}
2\delta_{2}=(\frac{q-3}{2},\underbrace{\frac{q-1}{2},\ldots}_{m-4},\frac{q-1}{2},\frac{q-3}{2},\frac{q+1}{2});\\
2\delta_{3}=(\frac{q-3}{2},\underbrace{\frac{q-1}{2},\ldots}_{m-4},\frac{q-3}{2},\frac{q+1}{2},\frac{q-1}{2}).
\end{aligned} 
\right.$
\end{center}
Let the $q-$adic expansion of $2a$ be $2a=\sum_{i=0}^{m-1}a_{i}q^{i}$, then $a_{m-1}=\frac{q-3}{2},a_{m-2}=\cdots=a_{3}=\frac{q-1}{2}$ and $0\leq a_0,a_1,a_2\leq q-1$. Next we will prove that $a_0,a_1,a_2\in\{\frac{q-3}{2},\frac{q-1}{2},\frac{q+1}{2}\}$. It is easy to verify for $q=3$.
If $q\neq3$, we have the following.
\begin{itemize}
\item[1)]If there exists $i\in [0,2]$ with $\frac{q-5}{2}\geq a_{i}$, then $[2aq^{m-i-1}]_{2n}<(\frac{q-5}{2},\underbrace{q-1,\ldots}_{m-2},q-1)<2\delta_{3}<2a$, this leads to a contradiction with the assumption of $2a\in \MinRep_{2n}$.
\item[2)]If there exists $i\in [0,2]$ with $a_{i}\geq \frac{q+3}{2}$, then
\begin{center}
$\begin{cases}
[2aq^{m-i-1}]_{2n}> (\frac{q+3}{2},0,0,\ldots,0),    &\text{if}\ i\neq0;\\
[2aq^{m-i-1}]_{2n}>(\frac{q+1}{2},\frac{q+1}{2},0,\ldots,0),  &\text{if}\ i=0.
\end{cases}$
\end{center}  Note that $2n-[2aq^{m-i-1}]_{n}\in C_{2a}^{2n}$, but $$2n-[2aq^{m-i-1}]_{2n}<(\frac{q-3}{2},\frac{q-3}{2},q-1,\ldots,q-1)+2<2\delta_{3}+2\leq2a,$$ which leads to a contradiction with $2a\in \MinRep_{2n}$.
\end{itemize}
Note that $2\mid 2a$, $2\delta_3+2\leq 2a\leq2\delta_2-2$ and $a_0,a_1,a_2\in\{\frac{q-3}{2},\frac{q-1}{2},\frac{q+1}{2}\}$, then $(a_2,a_1,a_0)$ = $(\frac{q-1}{2},\frac{q-3}{2},\frac{q-3}{2})$, i.e., $2a=2\delta_2-2$. But $\delta_2-1$ is even, which leads to a contradiction with $2\nmid a$.
			
Summing up all the discussions above, we can conclude that there is no odd integer $a$ such that $a\in \MinRep_{n}$ for any $\delta_3+1\leq a\leq\delta_2-1$. This concludes our proof.
\end{proof}
\end{lemma}
\begin{lemma}\label{le13}
Let $q\equiv7\pmod{8}$, $m=2h+1\geq 5$ and $n=\frac{q^m+1}{2}$. Then the first three largest odd coset leaders modulo $n$ are:
$$\delta_{1}=\frac{q^{m}-1}{4}-\frac{q^{m-1}}{2},\ \delta_{2}=\frac{q^{m}+1}{4}-\frac{q^{m-1}+q^2}{2},\ 
\delta_{3}=\frac{q^{m}+1}{4}-\frac{q^{m-1}+q^3-q^2+q}{2}.$$ 
Moreover, $|C_{\delta_{1}}^{n}|=|C_{\delta_{2}}^{n}|=|C_{\delta_{3}}^{n}|=2m$.
\begin{proof}
Note that $q\equiv7\pmod{8}$ and $m=2h+1\geq 5$, then $q^m\equiv7\pmod8$ and $q^{m-1}\equiv1\pmod8$. Therefore $\delta_{2}^{'} $ is odd, $\delta_{1}^{'}$ and $\delta_{3}^{'}$ are even, where $\delta_{1}^{'}$, $\delta_{2}^{'}$, $\delta_{3}^{'}$ are given in Lemma \ref{Le5}. Hence, $\delta_1$ is the first largest odd coset leader. 
			
We claim that $\delta_{2},\delta_{3} \in \MinRep_{n}$, i.e., $2\delta_{2},2\delta_3\in \MinRep_{2n}$ by Lemma \ref{Le6}. It is easy to get that
\begin{center}
$[2\delta_{2}q^{k}]_{2n}$=$\begin{cases}
\frac{q^m+1}{2}+q^{k-1}-q^{k+2},& \text{if}\ 1\leq k\leq m-3;\\
\frac{q^m+1}{2}+q^{m-3}+1,& \text{if}\ k= m-2;\\
\frac{q^m+1}{2}+q^{m-2}+q,& \text{if}\ k=m-1.
\end{cases}$\end{center}\begin{center}
$[2\delta_{3}q^{k}]_{2n}$=$\begin{cases}
\frac{q^m+1}{2}+q^{k+2}+q^{k-1}-q^{k+3}-q^{k+1},& \text{if}\ 1\leq k\leq m-4;\\
\frac{q^m+1}{2}+q^{m-1}-(q^2-1)q^{m-4}+1,& \text{if}\ k= m-3;\\
\frac{q^m-1}{2}-(q^2-1)q^{m-3}+q,& \text{if}\ k=m-2;\\
\frac{q^m+1}{2}+q^{m-2}+q^2-q+1,& \text{if}\ k=m-1.
\end{cases}$
\end{center}
Therefore, we have $[2\delta_{2}q^{k}]_{2n}-2\delta_2,[2\delta_{3}q^{k}]_{2n}-2\delta_3>0$ and $q^m+1>[2\delta_{2}q^{k}]_{2n}+2\delta_2,[2\delta_{3}q^{k}]_{2n}+2\delta_3$ for any $k\in [1,m-1]$. Using the same method as in Lemma \ref{le12}, we have $\delta_2,\delta_3\in \MinRep_{n}$, $|C_{\delta_2}^{n}|=|C_{\delta_3}^{n}|=2m$ and $\delta_2$ is the second largest odd coset leader.
			
Next, we will prove that $\delta_3$ is the third largest odd coset leader. Let $a\in\MinRep_{n}$ be odd and $\delta_3+1\leq a\leq\delta_2-1$, then $2a\in\MinRep_{2n}$ with $2\delta_3+2\leq 2a\leq2\delta_2-2$ by Lemma \ref{Le6}. Denote the $q-$adic expansion of $2a$ as $2a=\sum_{i=0}^{m-1}a_{i}q^{i}$, note that 
\begin{center}
	$\left\{\begin{aligned}
		&2\delta_{2}=(\frac{q-3}{2},\underbrace{\frac{q-1}{2},\ldots}_{m-4},\frac{q-3}{2},\frac{q-1}{2},\frac{q+1}{2});\\
		&2\delta_{3}=(\frac{q-3}{2},\underbrace{\frac{q-1}{2},\ldots}_{m-5},\frac{q-3}{2},\frac{q+1}{2},\frac{q-3}{2},\frac{q+1}{2}).
	\end{aligned} 
	\right.$
\end{center} then $a_{m-1}=\frac{q-3}{2},a_{m-2}=\cdots=a_{4}=\frac{q-1}{2}$ and $0\leq a_0,a_1,a_2,a_3\leq q-1$. Using a similar approach as in Lemma \ref{le12}, we get $a_0,a_1,a_2,a_3\in\{\frac{q-3}{2},\frac{q-1}{2},\frac{q+1}{2}\}$. 
\\Furthermore, $2\mid2a$, $2\nmid a$ and $2\delta_3+2\leq 2a\leq2\delta_2-2$, then 
$$2a=(\frac{q-3}{2},\underbrace{\frac{q-1}{2},\ldots}_{m-5},\frac{q-3}{2},\frac{q+1}{2},\frac{q+1}{2},\frac{q-3}{2}) \ or \ (\frac{q-3}{2},\underbrace{\frac{q-1}{2},\ldots}_{m-5},\frac{q-1}{2},\frac{q-3}{2},\frac{q-3}{2},\frac{q-1}{2})$$
If $2a=(\frac{q-3}{2},\frac{q-1}{2},\ldots,\frac{q-3}{2},\frac{q+1}{2},\frac{q+1}{2},\frac{q-3}{2})$, note that $2n-[2aq^{m-k-1}]_{n}\in C_{2a}^{2n}$ for any $k\in [1,m-1]$ and $$2n-[2aq^{m-3}]_{2n}=(\frac{q-3}{2},\frac{q-3}{2},\frac{q+3}{2},\frac{q-3}{2},\underbrace{\frac{q-1}{2},\ldots}_{m-5},\frac{q-1}{2})<2a,$$ which leads to a contradiction to the assumption that $2a\in \MinRep_{2n}$.
\\If $2a=(\frac{q-3}{2},\frac{q-1}{2},\ldots,\frac{q-1}{2},\frac{q-3}{2},\frac{q-3}{2},\frac{q-1}{2})$, note that $$[2aq^{m-3}]_{2n}=(\frac{q-3}{2},\frac{q-3}{2},\frac{q-3}{2},\frac{q+1}{2},\underbrace{\frac{q-1}{2},\ldots}_{m-5},\frac{q+1}{2})<2a,$$ which leads to a contradiction with $2a\in \MinRep_{2n}$. 
			
Summing up all the discussions above, then there is no odd integer $a$ such that $a\in \MinRep_{n}$ for any $\delta_3+1\leq a\leq\delta_2-1$. This concludes our proof.
\end{proof}
\end{lemma}
\begin{theorem}\label{t13}
Let $m=2h+1\geq 5$ and $n=\frac{q^m+1}{4}$, then
\begin{itemize}
\item[(1)]If $q\equiv3\pmod8$, then the code $\C_{(n,q,\delta)}$ has parameters
$$\begin{cases}
[n,1,d\geq \delta_{1}],&\ \text{if}\ \ \frac{\delta_{2}+3}{2}\leq \delta\leq \frac{\delta_{1}+1}{2};\\
[n,2m+1,d\geq \delta_{2}],&\ \text{if}\ \frac{\delta_{3}+3}{2}\leq \delta\leq \frac{\delta_{2}+1}{2},
\end{cases}$$ $\delta_1$, $\delta_2$, $\delta_3$ are given in Lemma \ref{le12}.	
\item[(2)]If $q\equiv7\pmod8$, then the code $\C_{(n,q,\delta)}$ has parameters $[n,2mi,d\geq \delta_{i}]$ for $\frac{\delta_{i+1}+3}{2}\leq \delta\leq \frac{\delta_{i}+1}{2}$ $(i = 1,2)$, $\delta_i$ is given in Lemma \ref{le13}.
\end{itemize}
\begin{proof}
Similar to Theorem \ref{t7}, we achieve the results using Lemmas \ref{le12} and \ref{le13}.
\end{proof}
\end{theorem}
\begin{example}
Based on Theorem \ref{t13}, let $(q,m)=(3,5)$, the code $\C_{(n,q,10)}$ has parameters $[61,11,31]$, the code is distance-optimal code according to the Database of known best codes in (\cite{RefJ11})..
\end{example}
\section{Conclusions}\label{set6} The main contributions are as follows:
\begin{itemize}
\item Based on the discussions of odd coset leaders in some ranges, we obtain the parameters of narrow-sense negacyclic BCH code $\C_{(n,q,\delta)}$ with large dimension, where $n=\frac{q^m-1}{4},\frac{q^m+1}{4}$ (see Theorems \ref{t1}, \ref{t2}, \ref{t4}, \ref{t5}, \ref{t6}, \ref{t11}, \ref{t12}).
		
\item By determining the first three largest odd coset leaders modulo $\frac{q^m-1}{2}$ and $\frac{q^m+1}{2}$, we present the parameters of narrow-sense negacyclic BCH code $\C_{(n,q,\delta)}$ with small dimension, where $n=\frac{q^m-1}{4},\frac{q^m+1}{4}$ (see Theorems \ref{t7}, \ref{t13}). Furthermore, the weight distribution of neagcyclic BCH codes of length $n=\frac{q^m-1}{4}$ is given for $\delta$ in some ranges (see Theorems \ref{t8}, \ref{t9}, \ref{t10}).
\end{itemize}
\section*{Acknowledgements}
The work was supported by the National Natural Science Foundation of China (Nos.12271137,12271335,\\62201009) and Natural Science Foundation of Anhui Province (No.2108085QA06).

\section*{Declarations}
\noindent{\textbf{Ethics approval and consent to\  participate} They have no known competing financial interests or personal relationships that might influence the work reported herein. All authors gave their informed consent.
	
\noindent{\textbf{Competing interests} The authors have declared that no completing interests exist.
		
\noindent{\textbf{Conflict of Interests} There is no conflict of interest.

\end{document}